\documentclass[useAMS,usenatbib,referee]{mn2e}

\usepackage{lscape}
\usepackage{graphicx}
\usepackage{txfonts}
\usepackage{longtable}

\title[Return of Activity in Comet 133P/Elst-Pizarro]{The Return of Activity in Main-Belt Comet 133P/Elst-Pizarro}

\author[H. H. Hsieh, D. Jewitt, P. Lacerda, S. C. Lowry, and C. Snodgrass]
       {Henry H. Hsieh$^{1}$\thanks{Email: h.hsieh@qub.ac.uk},
        David Jewitt$^{2}$, Pedro Lacerda$^{1,3}$, Stephen C. Lowry$^{4}$, and Colin Snodgrass$^{5,6}$
	\thanks{Some of the presented data were obtained at the W. M.
            Keck Observatory, which is operated as a scientific partnership
            among the California Institute of Technology, the University of
            California, and the National Aeronautics and Space Administration.
            The Observatory was made possible by the generous financial support
	    of the W. M. Keck Foundation.  Some data were also obtained at
	    Small and Moderate Aperture Research Telescope System (SMARTS) facilities
            at Cerro Tololo under programme NOAO-07B-0506, and at
	    European Southern Observatory (ESO) facilities at La Silla under
	    programmes 079.C-0384A and 081.C-0822A.
        }\\
        $^1$Astrophysics Research Centre, Queen's University,
         Belfast, BT7 1NN, United Kingdom\\
        $^2$Dept. of Earth and Space Sciences and Institute for Geophysics and Planetary Physics, UCLA, 3713 Geology Building, Box 951567,
	 Los Angeles, CA 90095, USA\\
        $^3$Newton Fellow\\
        $^4$Centre for Astrophysics and Planetary Science, University of Kent,
         Canterbury, CT2 7NH, United Kingdom\\
        $^5$Max Planck Institute for Solar System Research, Max-Planck-Str. 2, 37191 Katlenburg-Lindau, Germany\\
        $^6$European Southern Observatory, Alonso de Cordova 3107, Casilla
         19001, Santiago 19, Chile}
\begin{document}

\date{Submitted, 2009-10-27; Accepted, 2009-11-25}

\pagerange{\pageref{firstpage}--\pageref{lastpage}} \pubyear{2009}

\maketitle

\label{firstpage}

\begin{abstract}
Comet 133P/Elst-Pizarro is the first-known and currently best-characterised
member of the main-belt comets, a recently-identified class of objects
that exhibit cometary activity but which are dynamically indistinguishable from
main-belt asteroids.  We report here on the results of a multi-year
monitoring campaign from 2003 to 2008, and
present observations of the return of activity in 2007.  We find a pattern
of activity consistent with the seasonal activity modulation hypothesis
proposed by Hsieh et al. (2004, AJ, 127, 2997).  Additionally, recomputation
of phase function parameters using data in which 133P
was inactive yields new IAU parameters of
$H_R=15.49\pm0.05$~mag and $G_R=0.04\pm0.05$, and linear parameters of
$m_R(1,1,0)=15.80\pm0.05$~mag and $\beta=0.041\pm0.005$~mag~deg$^{-1}$.
Comparison between predicted magnitudes using these new parameters and the
comet's actual brightnesses during its 2002 and 2007 active periods reveals
the presence of unresolved coma during both episodes, on the order of
$\sim$0.20 of the nucleus cross-section in 2002 and $\sim$0.25 in 2007.
Multifilter observations during 133P's 2007 active outburst yield mean nucleus
colours of $B-V=0.65\pm0.03$, $V-R=0.36\pm0.01$, and $R-I=0.32\pm0.01$, with no
indication of significant rotational variation, and similar colours for the trail.
Finally, while 133P's trail appears shorter and weaker in 2007 than in 2002, other
measures of activity strength such as dust velocity and coma contamination of
nucleus photometry are found to remain approximately constant.  We attribute changes
in trail strength to the timing of observations and projection effects, thus
finding no evidence of any substantial decrease in activity strength between 2002
and 2007.

\end{abstract}


\section{INTRODUCTION}
Discovered on 1996 August 7 \citep{els96}, Comet 133P/Elst-Pizarro (also
designated 7968 Elst-Pizarro; hereafter 133P) orbits
in the main asteroid belt ($a=3.156$~AU, $e=0.165$,
$i=1.39^{\circ}$).  It has a Tisserand parameter (with respect to
Jupiter) of $T_J=3.184$, while classical comets have $T_J<3$
\citep{vag73,kre80}.
In 2005, two more objects displaying cometary activity that are likewise
dynamically indistinguishable from main-belt asteroids were identified:
P/2005 U1 (Read) \citep{rea05} and 176P/LINEAR (also known as asteroid
118401 (1999 RE$_{70}$))
\citep{hsi06c}.  Their discoveries led to the designation of a new
cometary class --- the main-belt comets (MBCs) --- among which 133P is also
classified \citep{hsi06b}.  A fourth MBC, P/2008 R1 (Garradd), has also
since been discovered \citep{gar08,jew09}.

Despite the initial excitement over the discovery of the cometary nature of
133P in 1996, no physical studies or monitoring reports were published in
the refereed literature until the comet's activity was re-observed in 2002
\citep{hsi04,low05}.  Consequently, little is known about 133P's active behaviour
in that intervening period.  Since knowledge of the timing of
active episodes can constrain hypotheses concerning the
source of the activity, we report the results of our own monitoring campaign,
which began following 133P's active outburst in 2002 and which culminated
in observations of renewed activity in 133P in 2007.

\section{OBSERVATIONS\label{observations}}

Since 133P's 2002 active episode, we have monitored the comet for evidence of
recurrent dust emission using the University of Hawaii (UH) 2.2-m telescope
and the 10-m Keck I telescope, both on Mauna Kea, the 1.3-m telescope
operated by the Small and Moderate Aperture Research Telescope System
(SMARTS) Consortium at Cerro Tololo, and the 3.58-m New Technology Telescope
(NTT) operated by the European Southern Observatory (ESO) at La Silla.  All
observations reported here were obtained under photometric conditions.
Details of these monitoring observations are listed in Table~\ref{obs_elstpiz}.

Observations with the UH 2.2-m telescope were made using a Tektronix
2048$\times$2048 pixel CCD with an image scale of $0\farcs219$ pixel$^{-1}$
behind Kron-Cousins BVRI filters.
Observations with Keck were made using the Low Resolution Imaging
Spectrometer (LRIS) imager \citep{oke95}
which employs a Tektronix 2048$\times$2048 CCD with an image scale of
$0\farcs210$~pixel$^{-1}$ and Kron-Cousins BVRI filters.
Observations with the SMARTS 1.3-m were made using the optical channel of
A Novel Double-Imaging Camera (ANDICAM) which employs a Fairchild 447
2048$\times$2048 CCD with an image scale of $0\farcs369$~pixel$^{-1}$
(using 2$\times$2 binning) and Johnson-Kron-Cousins BVRI filters.
Observations with the NTT in 2007 were made using the ESO
Multi-Mode Instrument (EMMI) \citep{dek86} which employs two adjacent
2048$\times$4096 MIT/LL CCDs with image scales of $0\farcs332$~pixel$^{-1}$
(using 2$\times$2 binning) and Bessel BVRI filters, while observations in
2008 were made using the ESO Faint Object
Spectrograph and Camera (EFOSC2) \citep{buz84} which employs a Loral/Lesser
2048$\times$2048 CCD with an image scale of $0\farcs24$~pixel$^{-1}$
(using 2$\times$2 binning) and Bessel BVR and Gunn i filters.
Except for those conducted with the SMARTS 1.3-m telescope, all
observations were made while tracking our target non-sidereally to
prevent trailing of the object.
For SMARTS 1.3-m observations, non-sidereal tracking was
not available, and as such, exposure times were selected such that the
trailing of the object during the course of a single exposure would be less
than $0\farcs5$, well below the typical full width at half-maximum (FWHM) seeing at the 1.3-m site.

Standard image preparation (bias subtraction and flat-field reduction) was
performed for all images.  Flat fields were constructed from dithered images
of the twilight sky.  Photometry of \citet{lan92} standard stars and field
stars was obtained by measuring net fluxes (over sky background) within
circular apertures, with background sampled from surrounding circular annuli.
Comet photometry was performed using circular apertures of different radii
(ranging from $2\farcs0$ to $5\farcs0$), but to avoid the contaminating
effects of the coma, background sky statistics were measured manually in
regions of blank sky near, but not adjacent, to the object.  Several
(5--10) field stars in the comet images were also measured to correct for
minor extinction variations during each night.

\section{RESULTS \& DISCUSSION\label{results}}

\subsection{Monitoring Campaign\label{monitoring}}

\begin{figure}
\includegraphics[width=6.5in]{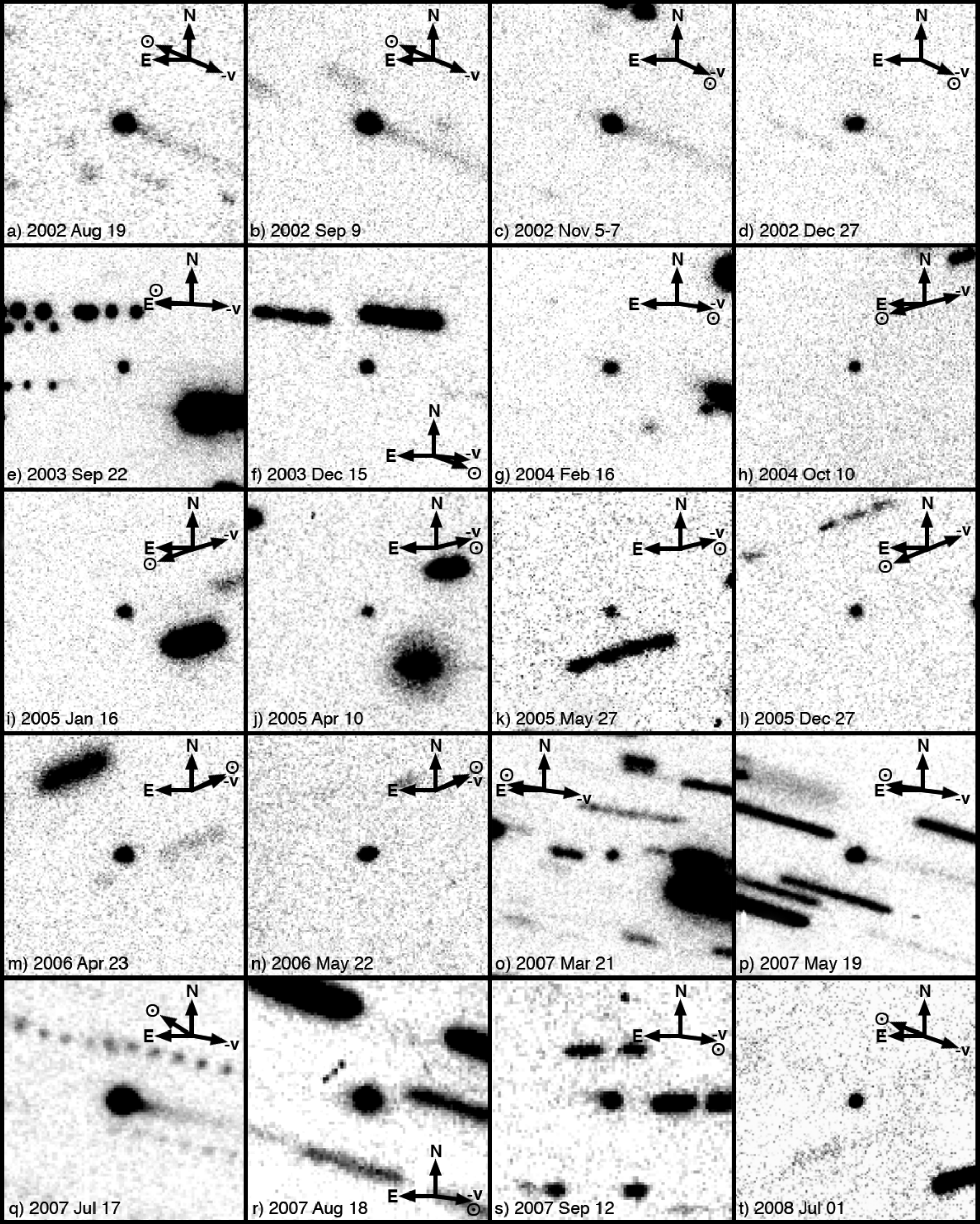}
\caption{\small Composite images of 133P, from $R$-band images taken during
  observations detailed in Table~\ref{obs_elstpiz}.  Each image is
  $0\farcm5\times0\farcm5$ with 133P at the center and arrows indicating
  North (N), East (E), the negative heliocentric velocity vector ($-$v),
  and the direction toward the Sun ($\odot$).
  Composite images are constructed from UH 2.2-m data
  unless otherwise stated and comprise total effective exposure times of
  (a) 2300~s, (b) 2100~s, (c) 4200~s, (d) 4200~s,
  (e) 500~s  (on Keck, equivalent to 10330~s on the UH 2.2-m),
  (f) 900~s,
  (g) 300~s  (on Keck, equivalent to 6200~s on the UH 2.2-m),
  (h) 200~s  (on Keck, equivalent to 4130~s on the UH 2.2-m),
  (i) 600~s, (j) 1200~s, (k) 1200~s, (l) 1200~s,
  (m) 1200~s, (n) 1200~s, (o) 1200~s, (p) 3300~s,
  (q) 900~s  (on the NTT, equivalent to 2380~s on the UH 2.2-m),
  (r) 4200~s (on the CT 1.3-m, equivalent to 1470~s on the UH 2.2-m),
  (s) 4140~s (on the CT 1.3-m, equivalent to 1450~s on the UH 2.2-m), and
  (t) 600~s  (on the NTT, equivalent to 1590~s on the UH 2.2-m).
  Data in panels (a) through (e) were previously presented in
  \citet{hsi04}
  }
\label{images_133p}
\end{figure}

For all monitoring observations, individual $R$-band images (aligned on the
object's photocenter using linear interpolation) from each night were
combined into single composite images (Fig.~\ref{images_133p}).  For
reference, we also show composite images from
133P's 2002 active phase
\citep[Figs.~\ref{images_133p}a--\ref{images_133p}d;][]{hsi04}.
Activity is marginally visible in images from 2007 May 19,
2007 August 18, and 2007 September 12
(Figs.~\ref{images_133p}p, \ref{images_133p}r, \ref{images_133p}s),
while the comet's characteristic dust trail is clearly
visible in the image from 2007 July 17 (Fig.~\ref{images_133p}q).
We find no evidence of activity in images from 2003 September 22
through 2007 March 21 (Figs.~\ref{images_133p}e--\ref{images_133p}o)
and from 2008 July 1 (Fig.~\ref{images_133p}t).
In all images, even those obtained while 133P was active, the
FWHM of the object's surface brightness profile is consistent with the typical
FWHM seeing at the time of night when those images were obtained, implying
that little or no coma is present.

\begin{figure}
\includegraphics[width=6.5in]{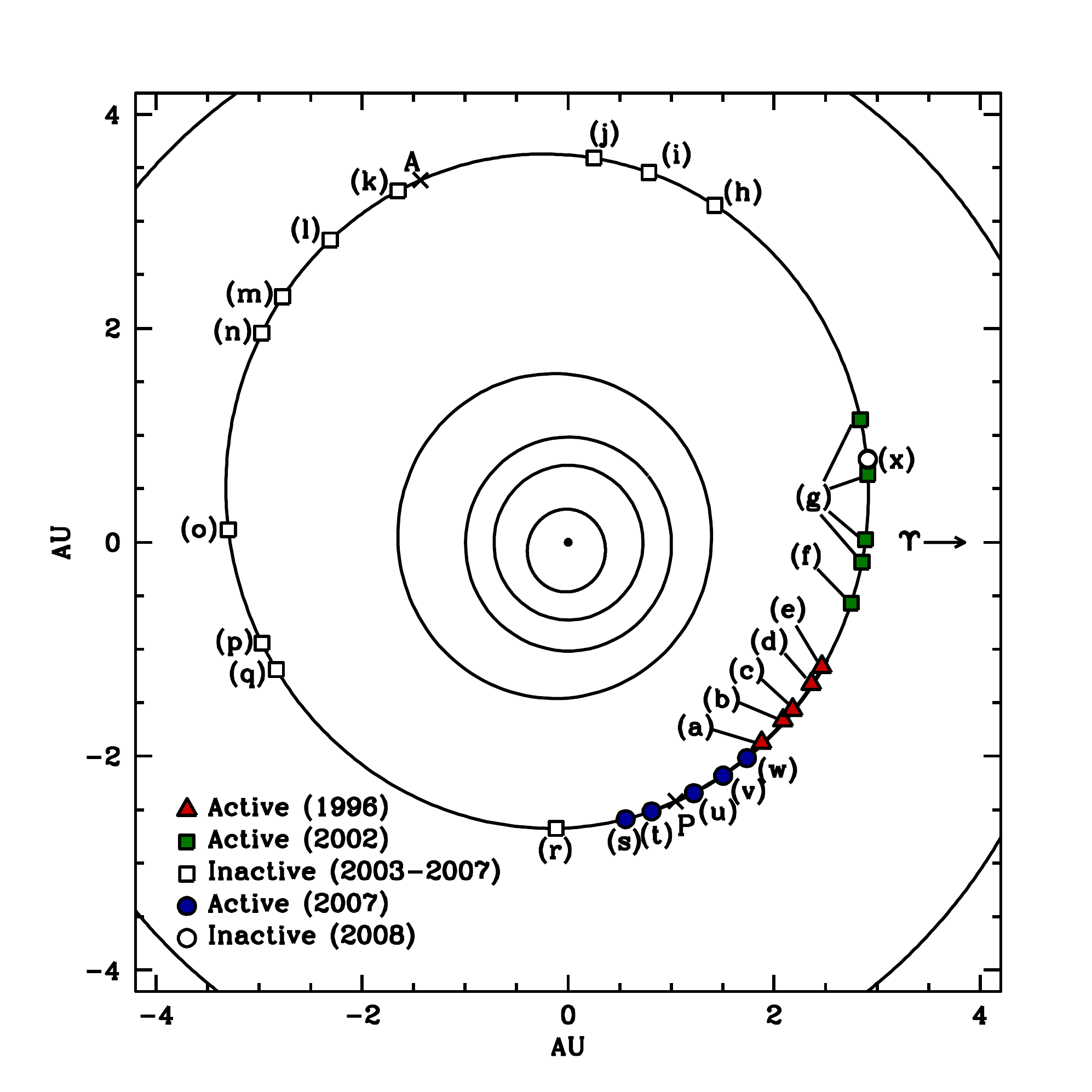}
\caption{\small Orbital position plot of reported 133P observations detailed
  in Table~\ref{obs_elstpiz} and in \citet{hsi04}.  The Sun is shown at the
  center as a solid dot, with the orbits of Mercury, Venus, Earth, Mars, 133P,
  and Jupiter (from the center of the plot outwards) are shown as black lines.
  Solid triangles mark positions where 133P was observed to be active in 1996,
  while solid squares mark active positions in 2002, and solid circles mark
  active positions in 2007.  Open squares mark positions where no activity
  was detected in 133P between 2003 and 2007, while the open circle marks
  the position where 133P was observed to be inactive in 2008.
  Perihelion (P) and aphelion (A) positions are also marked. References:
  (a) 1996 Jul 14 \citep{els96};
  (b) 1996 Aug 09 \citep{els96};
  (c) 1996 Aug 21-22 \citep{els96,pra96};
  (d) 1996 Sep 16, 18 \citep{boe96,ham96};
  (e) 1996 Oct 04 \citep{boe96};
  (f) 2002 Jul 13 \citep{low05};
  (g) 2002 Aug 19, 2002 Sep 07-09, 2002 Nov 05-07,
       and 2002 Dec 27-28 \citep{hsi04}
  (h) 2003 Sep 22 \citep{hsi04};
  (i) 2003 Dec 13-15;
  (j) 2004 Feb 16;
  (k) 2004 Oct 10;
  (l) 2005 Jan 16;
  (m) 2005 Apr 10;
  (n) 2005 May 27-28;
  (o) 2005 Dec 27;
  (p) 2006 Apr 23;
  (q) 2006 May 22-25;
  (r) 2007 Mar 21;
  (s) 2007 May 19;
  (t) 2007 Jun 11 \citep{jew07};
  (u) 2007 Jul 17-20,
  (v) 2007 Aug 18;
  (w) 2007 Sep 12;
  (x) 2008 Jul 01,
  where (i) -- (s) and (u) -- (x) are from this work.}
\label{actv133p}
\end{figure}

In Figure~\ref{actv133p}, we mark the positions where we observed 133P to be
active or where others reported it to be active,
as well as positions where we observed it to be inactive,
on a plan view of its orbit.
The figure shows that reports of activity in 133P are approximately confined to
the quadrant following perihelion, with the earliest detection of activity
occurring shortly before perihelion at a true anomaly of
$\nu\approx350^{\circ}$
and the latest detection occurring at $\nu\approx90^{\circ}$.  
This activity profile is consistent with the hypothesis of seasonal activity modulation
described in \citet{hsi04} and \citet{hsi06a}, whereby 133P's activity is driven by the
sublimation of a localised patch of exposed volatile material confined to either
the ``northern'' or ``southern'' hemisphere of the body.
Assuming non-zero obliquity, activity then only occurs during
the portion of the orbit
when that active site receives enough solar heating to drive
sublimation, i.e., during that hemisphere's ``summer''.
We note that our observations
of 133P on 2008 July 1 at the NTT showed it to be inactive despite the object
being observed to be active at nearly the same orbital position in 2002.
We attribute this discrepancy to a combination of the low
signal-to-noise of this observation and the expected extremely weak
activity of 133P at that point in its orbit \citep{hsi04}.

\subsection{Photometric Activity Detection and Measurement\label{activitydetection}}

When no coma is clearly
visible for an object, an alternate method for detecting activity is
examination of its photometric behaviour: i.e., determining whether it is consistent
with an inactive object of a fixed size, or whether it shows anomalous brightening
over a certain portion of its orbit.  This type of
analysis led to the discovery of activity in 95P/(2060) Chiron
\citep{tho88,bus88,mee89,har90}.  In applying this approach to
133P, we recall that \citet{hsi04} originally derived linear and IAU $H$,$G$
phase function solutions for 133P using data taken
in 2002 when the object was visibly emitting dust.  In the case of that
data set, 133P's activity was judged to contribute negligibly to
nucleus photometry (as no significant coma was detected)
and was thus assumed to affect phase function derivations similarly negligibly.
Having since accumulated a substantial set of observations while 133P was
entirely inactive, though, we can now assess the validity of this
neglect by deriving new phase function solutions and comparing the results
to those of \citet{hsi04}.

We caution that, unlike the data used by \citet{hsi04},
the photometric data used in this follow-up analysis (2003 Sep 22 to 2007 Mar 21) all
consist of ``snapshot observations'', which are short sequences of exposures
at unknown rotational phases, instead of full lightcurves.  This
caveat is significant because
rotation of the body is expected to cause deviations in measured
brightness by as much as 0.2~mag from the comet's
true mean brightness at a given time
\citep{hsi04}.  Given a sufficiently large data set, however, we expect that
the average of these fluctuations will approach zero, allowing us to derive
reasonably accurate phase function solutions without necessarily knowing the rotational
phase at which each individual photometry point was obtained.
Nonetheless, the lack of rotational phase information for our snapshot
observations remains a source of uncertainty.

\begin{figure}
\includegraphics[width=6.5in]{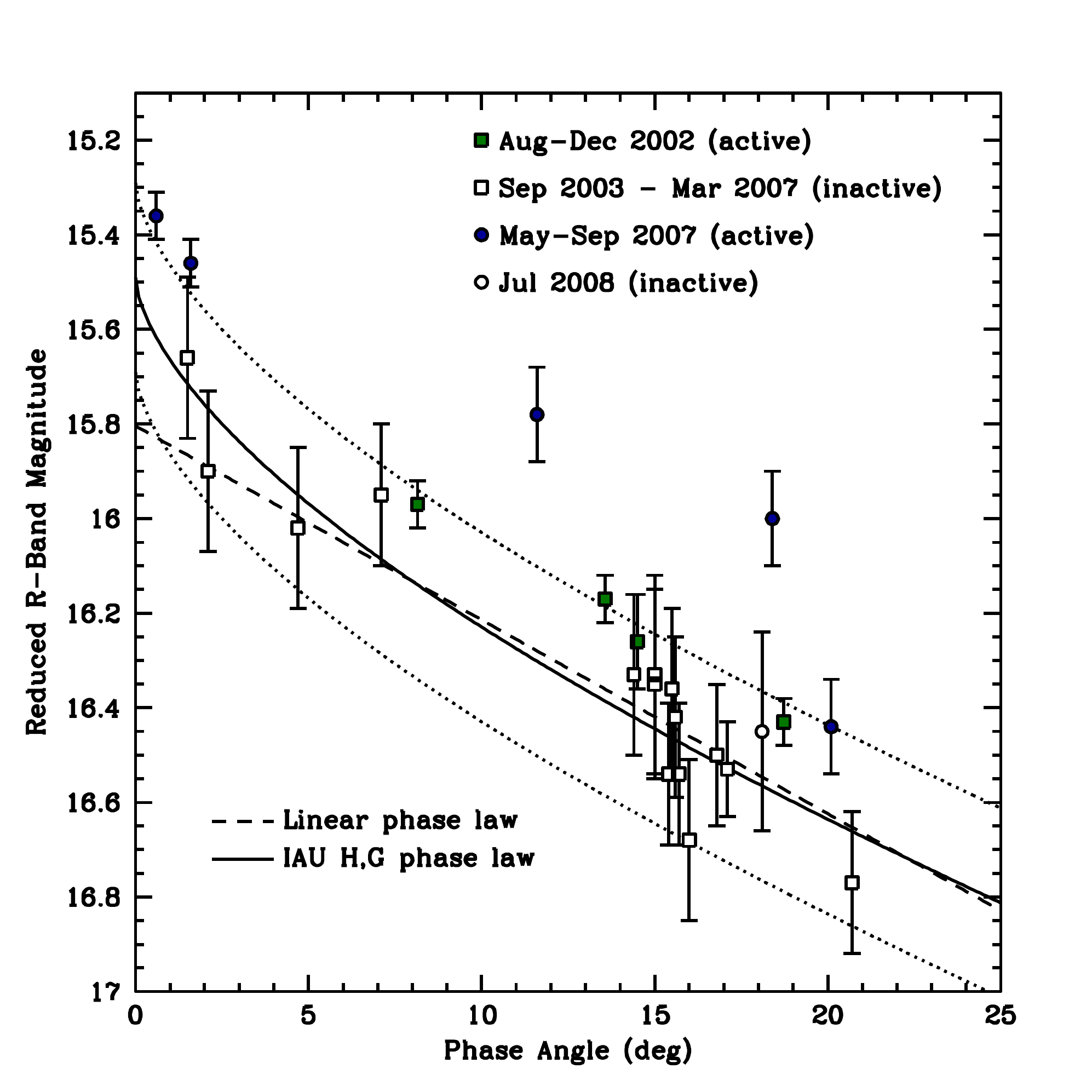}
\caption{\small Phase functions for 133P.  Points are estimated $R$-band magnitudes
  (normalised to heliocentric and geocentric distances of 1 AU; tabulated in
  Table~\ref{obs_elstpiz}) at the midpoint of the full photometric range of
  the nucleus's rotational lightcurve.  Solid symbols denote photometry
  obtained while 133P was visibly active, while open symbols denote
  photometry obtained while 133P appeared to be inactive.
  The dashed line represents a least-squares fit
  (excluding photometry points for which $\alpha<5^{\circ}$ where an opposition surge effect
  is expected) to a linear phase function
  where $m_R(1,1,0)=15.80\pm0.07$~mag and $\beta=0.041\pm0.005$~mag~deg$^{-1}$.
  The solid line represents an IAU ($H,G$) phase function fit where
  $H_R=15.49\pm0.05$~mag and $G_R=0.04\pm0.05$, while the dotted lines
  indicate the expected range of possible magnitude variations ($\sim$0.2~mag) due to
  the object's rotation.
}
\label{phaselaws}
\end{figure}

We compute the reduced magnitude, $m_R(1,1,\alpha)$, of 133P at the time
of each observation using
\begin{equation}
m_R(1,1,\alpha) = m_{mid}(R,\Delta,\alpha) - 5\log(R\Delta)
\end{equation}
where $R$ is the heliocentric distance of the object in AU, $\Delta$ is the
object's geocentric distance in AU, and $m_{mid}(R,\Delta,\alpha)$ is the estimated
$R$-band magnitude at the midpoint of the full photometric range
of the rotational lightcurve (Table~\ref{obs_elstpiz}).
For observations of full lightcurves, $m_{mid}$ is determined by simply
plotting the data and locating the midpoint between the maximum and minimum
values of the lightcurve.  For snapshot observations, $m_{mid}$ is
generally taken to be the mean of the available photometry data with large
error bars applied to reflect rotational phase uncertainties,
assuming a full possible photometric range of 0.40~mag.
We fit
reduced magnitude values to both a linear phase function and an IAU phase
function, finding best-fit values of $m_R(1,1,0)=15.80\pm0.07$~mag and
$\beta=0.041\pm0.005$~mag~deg$^{-1}$ for the linear phase function where
\begin{equation}
m_R(1,1,\alpha) = m_R(1,1,0) + \beta\alpha .
\end{equation}
We also find best-fit values of
$H_R=15.49\pm0.05$~mag and $G_R=0.04\pm0.05$ for the IAU phase function
as defined in \citet{bow89}.
Photometry obtained at phase angles of $\alpha<5^{\circ}$, where an opposition surge effect is
expected, are included in the derivation of the IAU phase function but omitted from the derivation
of the linear phase function.
We plot our best-fit solutions in Figure~\ref{phaselaws}.
A modest amount of scatter around our solutions is present, as expected, but
in all cases the deviations from the best-fit phase functions
are consistent with expected brightness fluctuations due to 133P's rotation.
Due to the uncertainty of the active status of 133P on 2008 July 01
(\S\ref{monitoring}), the photometry from that night is plotted but was not
included in the computation of the best-fit phase functions.

While the slope parameters of both newly-derived functions are consistent with
the parameters computed in \citet{hsi04}
($\beta=0.044\pm0.007$~mag~deg$^{-1}$; $G_R=0.026\pm0.1$), both newly-derived
absolute magnitudes are $\sim$0.2~mag fainter than their
previously derived values ($m_R(1,1,0)=15.61\pm0.01$~mag;
$H_R=15.3\pm0.1$~mag), strongly suggesting that the previously derived
parameters were affected by contamination by 133P's dust emission.
This contamination is assumed to consist of a combination of coma and
the portion of the dust trail (as projected in the plane of the sky)
contained within the seeing disc.
This suggestion of dust contamination is reinforced
by Figure~\ref{phaselaws} where we note that photometry from both
133P's 2002 and 2007 active phases are consistently brighter than expected
from our new phase function solutions.  Because most of the data points
from 2002 and 2007
are mean magnitudes derived from fully sampled lightcurves, brightness
fluctuations due to rotation cannot account for the discrepancies.

Assuming that the discrepancy between an observed magnitude, $m_{mid}$,
and expected magnitude, $m_{exp}$, is due to dust contamination,
the scattering surface area of the dust, $A_d$, is given by
\begin{equation}
A_d = A_n\left({A_d\over A_n}\right)
    = A_n\left(10^{0.4(m_{exp} - m_{mid})} - 1\right)
\label{eqadust}
\end{equation}
where $A_n=\pi r_e^2=1.13\times10^7$~m$^2$
is the scattering cross-section of the nucleus \citep{hsi09},
and albedos of the nucleus and dust are assumed to be equal.
Assuming optically thin dust, the total dust mass, $M_d$,
can then be estimated from
\begin{equation}
M_d \sim {4\over 3}\pi\rho a_d^3 \left({A_d\over\pi a_d^2}\right)
\label{eqmdust}
\end{equation}
where we adopt typical dust grain radii of $a_d=10$~$\mu$m and
a bulk grain density of $\rho_d=1300$~kg~m$^{-3}$ \citep[{\it cf}.][]{hsi04}.

For reference, we also compute $Af\rho$ \citep[{\it cf}.][]{ahe84} for each
set of observations where the parameter is given by
\begin{equation}
Af\rho = {(2R\Delta)^2\over \rho} 10^{0.4[m_{\odot}-m_R(R,\Delta,0)]}
\label{eqafrho}
\end{equation}
where $R$ is in AU, $\Delta$ is in cm, $\rho$ is the physical radius in cm of
a $4\farcs0$-radius photometry aperture at the distance of the comet, and
$m_R(R,\Delta,0)$ is the phase-angle-corrected $R$-band magnitude of the
comet measured using a $4\farcs0$-radius aperture, which we calculate using
\begin{equation}
m_R(R,\Delta,0) = m_{mid}(R,\Delta,\alpha)
                   + 2.5\log\left[(1-G)\Phi_1(\alpha)+G\Phi_2(\alpha)\right]
\end{equation}
where $\Phi_1$ and $\Phi_2$ are given by
\begin{equation}
\Phi_1=\exp\left[-3.33\left(\tan{\alpha\over 2}\right)^{0.63}\right]
\end{equation}
\begin{equation}
\Phi_2=\exp\left[-1.87\left(\tan{\alpha\over 2}\right)^{1.22}\right]
\end{equation}
\citep{bow89}.

Using Equations~\ref{eqadust}, \ref{eqmdust}, and \ref{eqafrho},
we compute $A_d$, $M_d$, and $Af\rho$
for each set of observations from 2002 and 2007
during which 133P was observed to be active, and tabulate the results in
Table~\ref{dustcontrib}. We find that for data from 2002, 
dust contamination is approximately constant
with a scattering surface area of
$\sim$0.20$A_n$ and a dust mass of
$M_d\sim4\times10^4$~kg contained within $\sim$3~arcsec ($\sim$4500~km
in August-November; $\sim$6500~km in December)
photometry apertures.
The relatively constant amount of dust over this
time period explains why we were able to derive reasonably accurate
slope parameters for 133P from our 2002 data despite arriving
at incorrect results for the comet's absolute magnitude due to the
dust contamination.

In data from July 2007, we find that 133P's inferred dust coma has a
strength comparable to that observed in 2002, having a scattering
surface area equivalent to $\sim$0.25$A_n$
and a dust mass of $M_d\sim5\times10^4$~kg contained within $\sim$4~arcsec 
($\sim$4700~km) photometry apertures.
The slightly larger amount of inferred dust in 2007 could indicate a higher rate
of dust production, but could also be due to
different viewing geometries, given that 133P was close to
opposition when observed in July 2007.  At this position, the antisolar
vector for 133P points very nearly directly behind the object as seen
from Earth, causing more of the dust trail to be
located within the seeing disc of the comet as projected on the sky.
Given the limitations of our observations, however, we are unable
to disentangle this possible projection effect from any intrinsic increase in
dust production.
Additionally, 133P's apparent brightness could also have been enhanced
by an opposition surge effect from the dust in its coma, though we
unfortunately lack observational constraints for quantifying this effect.

Given these various possible contributing factors to 133P's
enhanced brightness on 2007 July 17 and 20, we are unable to
determine whether 133P was more active on these dates compared to
2002 August 19 through 2002 November 7.  We can conclude, however, that
coma contamination is present in nucleus photometry performed for 133P
during both observing periods, and that the measured magnitude enhancements
suggest at least comparable levels of activity in each case.

The remainder of our photometry from 133P's 2007 active phase is
derived from incomplete lightcurve information, and as such, coma estimates
at these times have much larger uncertainties than at other times.
We find no definitive evidence of a coma on
2007 May 19, but find that the inferred comae on 2007 Aug 18 and
2007 Sep 12 are far stronger ($\sim$0.65$A_n$)
than in any other observations, a rather unexpected discovery given the minimal
amount of time elapsed since our 2007 July observations.  We suggest that the
large inferred dust contribution to nucleus photometry in August and September
could be at least partly due to geometric effects.  As can be seen in
Figures~\ref{images_133p}q-\ref{images_133p}s, the orientation of the
projection of the dust trail appears to change over this period of time.  
We caution that poor seeing during our August and September observations
and the small aperture (1.3~m) of the telescope used to obtain these data
mean that the observed morphology
(namely, the near-disappearance of the dust trail) cannot be considered entirely reliable.
If the observed morphology is believed, however,
much of the precipitous increase in 133P's apparent coma strength
between July and August could be due to the dust trail becoming almost directly aligned
behind the nucleus in August and September, thus becoming unavoidably included within our
photometry apertures.

We can account for this viewing geometry effect by integrating the scattering
surface area of the visible dust trail measured in July data (discussed below;
\S\ref{dusttrail}) and then assuming that it all falls within the photometry aperture
used to measure the nucleus magnitudes in August and September.
The net increase in dust scattering
surface area implied by photometry between 2007 July 20 and 2007 August 18
is $\sim$0.40$A_n$.
The integrated scattering surface area of the dust trail on 2007 July 20 over the
first 30~arcsec from the nucleus (the trail becomes too faint to measure reliably
beyond this point), however, is $\sim$0.20$A_n$, accounting for only about
half of the observed increase in dust contamination between
July and August.  The remainder of the observed increase could be partly due to
distant material in the dust trail
that was too diffuse to detect in trail form in
July data, but nevertheless contributed
positively to nucleus photometry when projected directly behind the nucleus in August and September.
It seems unlikely, however, that half of the dust in the trail could go undetected
in our July data, and as such, we surmise that at least part of the increase must
in fact be due to a real increase in dust production, which of course would certainly
be plausible at this early stage in 133P's active phase.

\subsection{The Lightcurve Revisited\label{ltcurverevisit}}

\subsubsection{Search for Rotational Colour Variations\label{colorvariations}}

\begin{figure}
\includegraphics[width=6.5in]{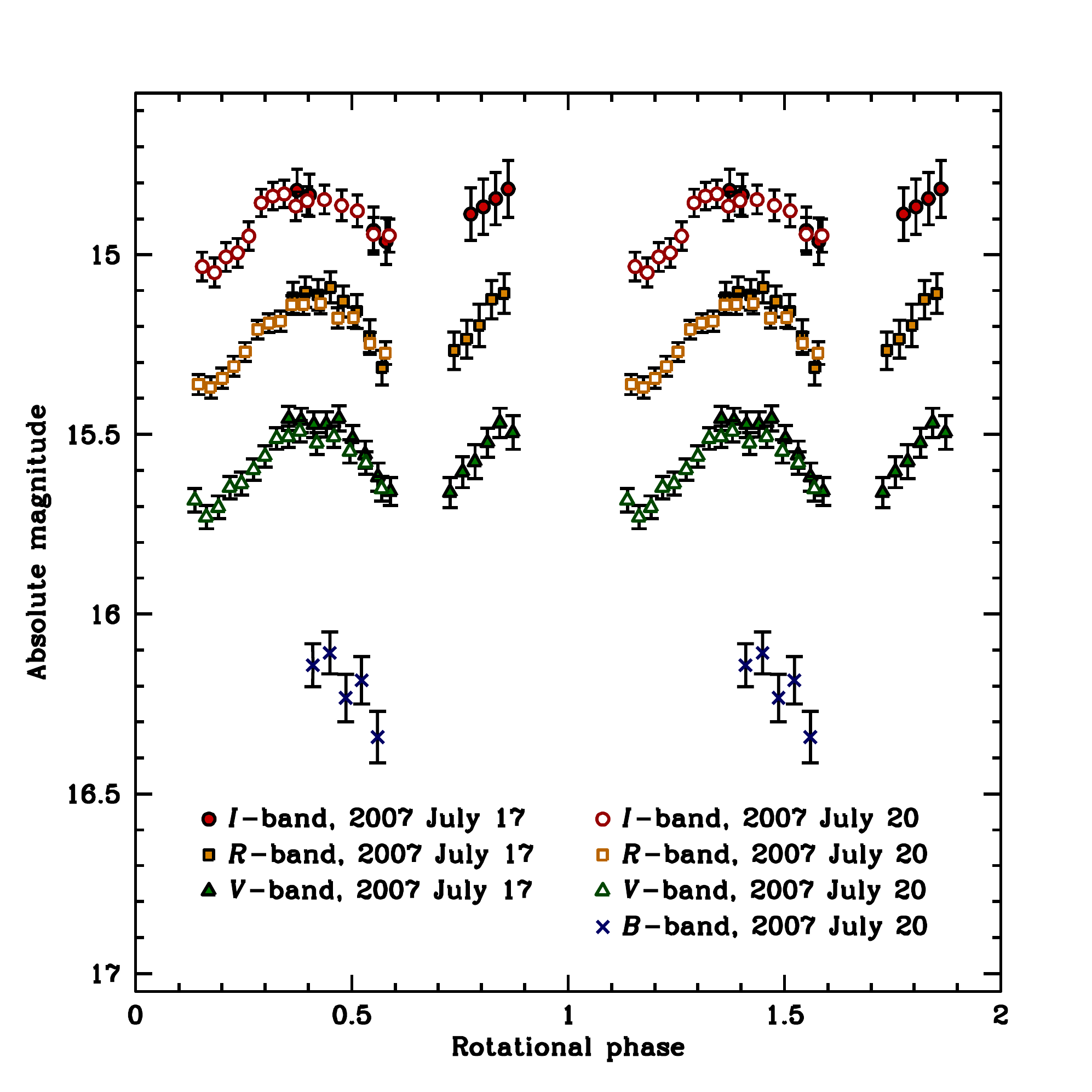}
\caption{\small Simultaneous $B$-band, $V$-band, $R$-band, and $I$-band lightcurves,
  phased to a rotation period of $P_{rot}=3.471$~hr,
  for 133P obtained using the NTT on 2007 July 17 and 2007 July 20.}
\label{nttltcurves}
\end{figure}

\begin{figure}
\includegraphics[width=6.5in]{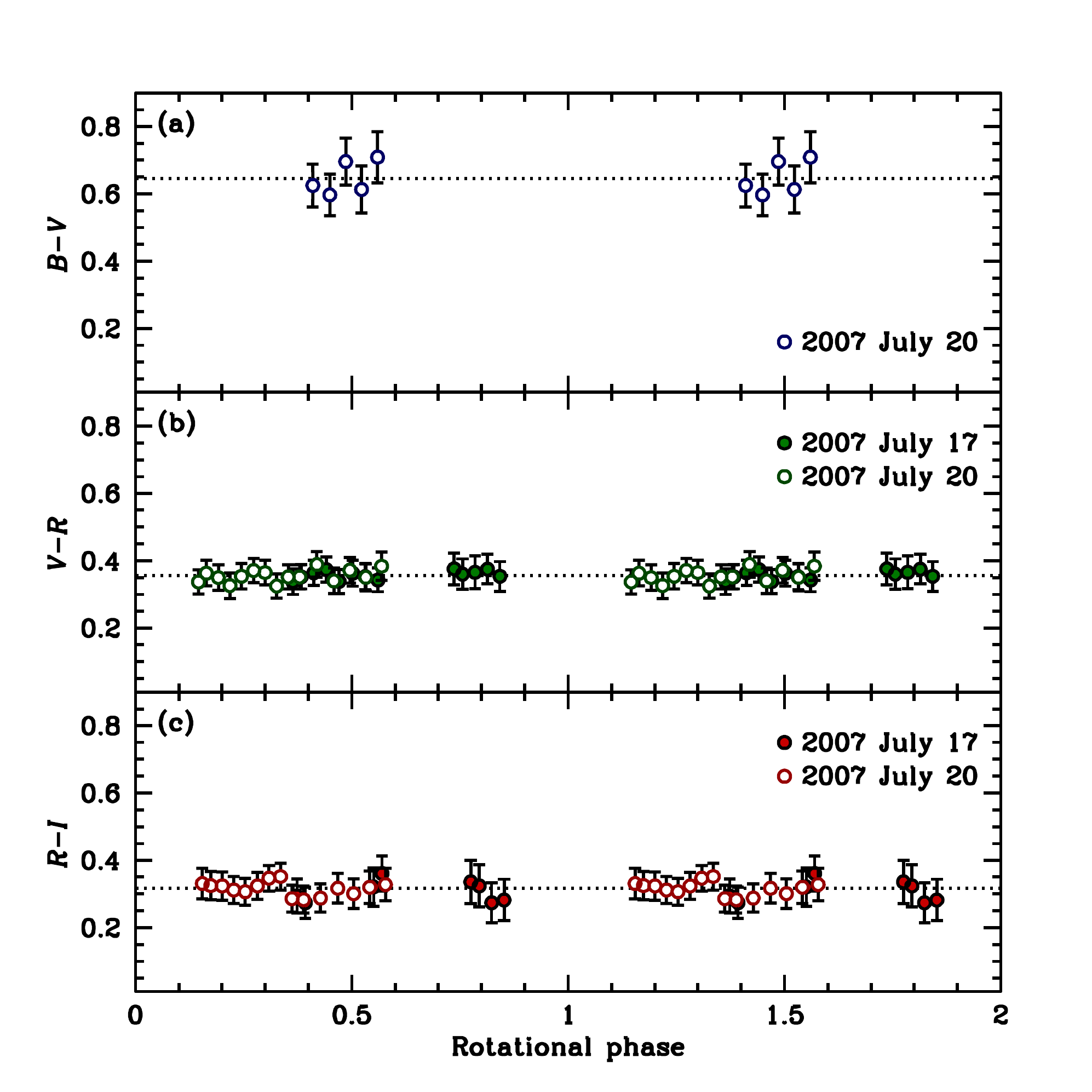}
\caption{\small Interpolated (a) $B-V$, (b) $V-R$, and (c) $R-I$ colours,
  phased to a rotation period of $P_{rot}=3.471$~hr, for 133P
  obtained using the NTT on 2007 July 17 and 2007 July 20}
\label{nttcolors}
\end{figure}

During our 2007 NTT run when 133P was active, we
observed the comet in continuously cycling filters ($VRI$ on 2007 July 17
and $BVRI$ on 2007 July 20).  Observations were made in this way
to allow us to obtain deep imaging of 133P in multiple filters and
also construct simultaneous lightcurves in each filter.
These lightcurves then allowed us to
search for surface colour inhomogeneities that, for example,
may constrain the position of the localised active site hypothesized by \citet{hsi04}.
These lightcurves, phased to a rotational period of $P_{rot}=3.471$~hr \citep{hsi04},
are plotted in Figure~\ref{nttltcurves}.
To then assess colour variation as a function of
rotational phase for each filter pair, we use linear interpolation to obtain the
magnitudes of the object in the second filter at times of observations in the
first filter and then plot the differences (Fig.~\ref{nttcolors}), again
phased to $P_{rot}=3.471$~hr.

We find mean nucleus colours of $B-V=0.65\pm0.03$~mag, $V-R=0.36\pm0.01$~mag, and
$R-I=0.32\pm0.01$~mag.  These values are somewhat different from the mean colours
found for 133P by \citet{hsi04}, but are within the range of individual values measured
in that work.  We regard the colour measurements presented here to be more
accurate since our repeated multifilter
observations of 133P here allowed us to account for both
rotational magnitude variations (via lightcurve interpolation)
and minor extinction variability (using field stars
as references for making differential photometric corrections).
The single sets of multifilter observations used to make 133P's previous colour measurements
did not permit either of these corrective measures.

Upon examining individual colour measurements, we find no conclusive evidence of
rotational colour inhomogeneity.  We find maximum colour variations of only
$\Delta(B-V)=0.11\pm0.13$~mag, $\Delta(V-R)=0.06\pm0.07$~mag, and
$\Delta(R-I)=0.08\pm0.08$~mag, where the non-systematic
distribution of even these small variations indicates that they are most likely due to
ordinary measurement uncertainties.  We note that this result does not rule out
the possibility that 133P's active area exhibits a different colour signature than
inactive surface material.  First, the coma that is likely present (\S\ref{activitydetection})
should act to obscure colour variations on the nucleus surface, with the precise amount
of obscuration varying with rotational phase as the ratio of the nucleus's scattering
cross-section to the coma's cross-section changes.

Furthermore, under the seasonal heating hypothesis \citep{hsi04,hsi06a},
the active site is in fact expected to
be illuminated by the Sun
at all rotational phases when near perihelion (assumed to be close to solstice)
when these observations were made.
The nucleus orientation at this time allows the active site to receive maximal solar
heating but also means that the active site is always in the line of sight as viewed
from Earth.
We suggest that more favourable conditions for detecting colour inhomogeneities will
occur around 133P's next pre-perihelion equinox (likely near $\nu\sim270^{\circ}$).  Based on prior observations
(\S\ref{monitoring}), the nucleus should be largely coma-free over this portion of
the orbit, and at equinox, the active site should pass into and out of the line of
sight as the nucleus rotates, maximising any colour variations.  We therefore
encourage additional rotationally-resolved colour measurements of 133P between late 2011
and early 2012.

\subsubsection{Implications for 133P's Pole Orientation}

\begin{figure}
\includegraphics[width=6.5in]{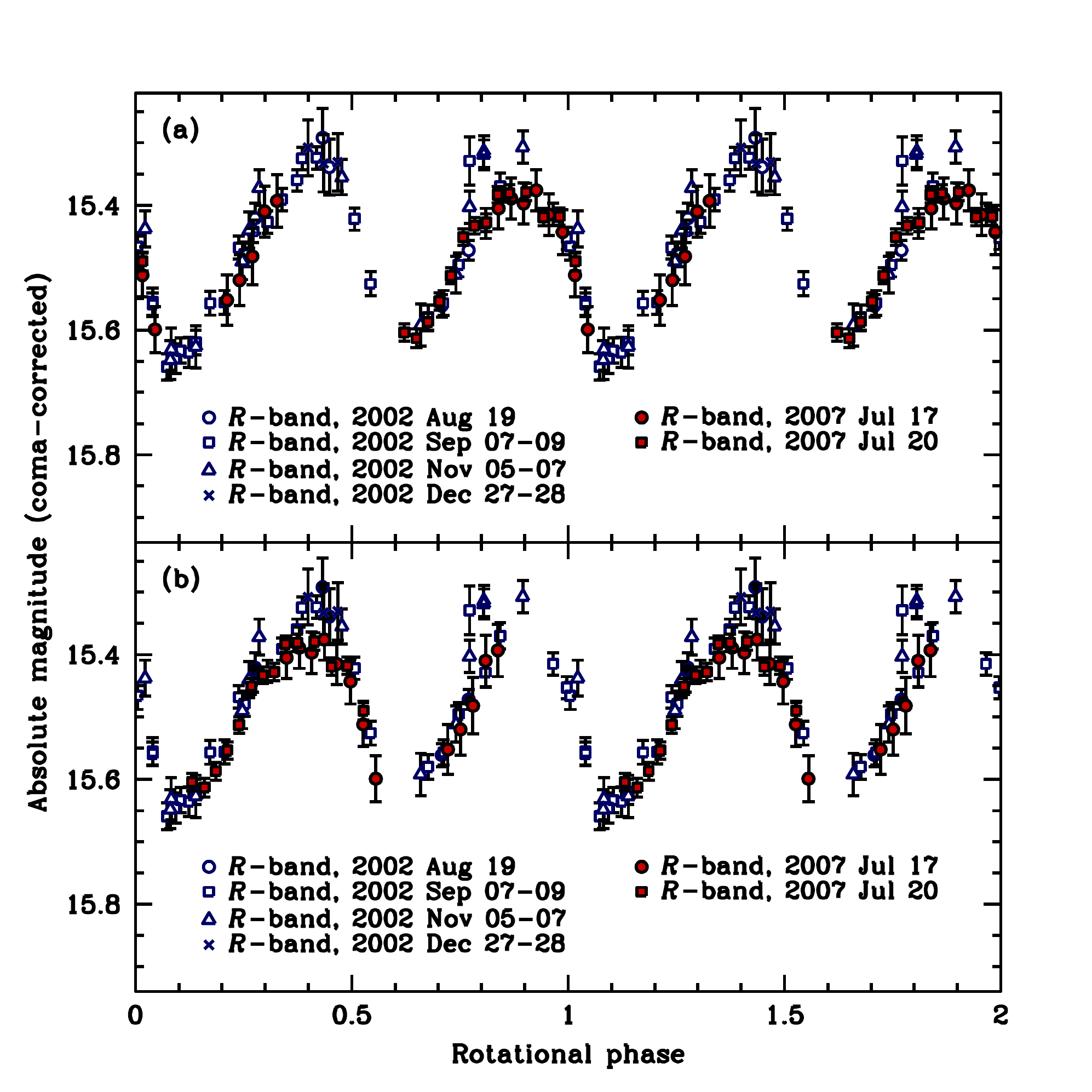}
\caption{\small $R$-band lightcurves
  for 133P obtained using the UH 2.2-m telescope in 2002 (open symbols) and the
  NTT in 2007 (solid symbols).  Data for each year are phased together to a
  rotation period of $P_{rot}=3.471$~hr, after which the two data sets from
  2002 and 2007 are aligned by eye.  Panels (a) and (b) show the two possible
  alignments of the 2002 and 2007 data.
}
\label{ltcurves0207}
\end{figure}

For reference, we remove the estimated dust contamination from both our 2002 and 2007
lightcurve data, and overplot the two sets of lightcurves (Fig.~\ref{ltcurves0207}).
Each of the two sets of data are phased self-consistently to $P_{rot}=3.471$~hr, though
given the great difficulty of phasing data together that are separated by almost 5 years to such a short
rotational period, the 2002 and 2007 data are simply aligned by eye.  Due to the
two-peaked nature of 133P's lightcurve, though, there is an ambiguity in performing
this alignment.  In one case (Fig.~\ref{ltcurves0207}a), the data can be aligned such
that the lightcurve shape and photometric range appear largely unchanged between the
two observation epochs.
In the second case (Fig.~\ref{ltcurves0207}b), the data can be aligned such that the
photometric range of lightcurve appears to decline to $\Delta m_R\sim0.25$~mag
in 2007 from $\Delta m_R\sim0.35$~mag in 2002.  In the latter case, it should be recalled that
the coma contribution to the data plotted has already been subtracted,
and as such the change in photometric range cannot be attributed to differences in the amount of coma.
Unfortunately, due to the incomplete sampling of the lightcurve in 2007, it is not possible
to resolve the ambiguity between these two cases.

This ambiguity is significant because of the implications of photometric range
behaviour for the orientation of 133P's rotational pole.
To gain more insight as to how the photometric range of 133P should change
depending on pole orientation and observing geometry, we
simulate its lightcurve behaviour using the model presented in \citet{lac07}.
We assume a simple prolate ellipsoidal shape for
the nucleus of 133P and render it at various observing geometries and rotational
phases.  At each rotational phase, the
light reflected back to the observer is integrated to generate lightcurve points.
The 2002 September coma-corrected photometric range for 133P was measured to be
$\Delta m_R=0.35$~mag, and so we use a nucleus axis ratio of $a/b=10^{0.4\Delta m_R}=1.39$
(it should be noted that this is a lower limit due to the unknown projection
angle at the time). We use a Lommel-Seeliger ``lunar''
scattering function \citep[{\it cf}.][]{fai05} which has no free parameters and
is appropriate for simulating the low
albedo \citep[$p_R=0.05\pm0.02$;][]{hsi09} surface of 133P.  To
simplify the geometry, we neglect the small orbital inclination
($i=1.4\degr$) of 133P and assume that it is coplanar with the Earth
(i.e., $i=0\degr$).

\begin{figure}
\includegraphics[width=6.5in]{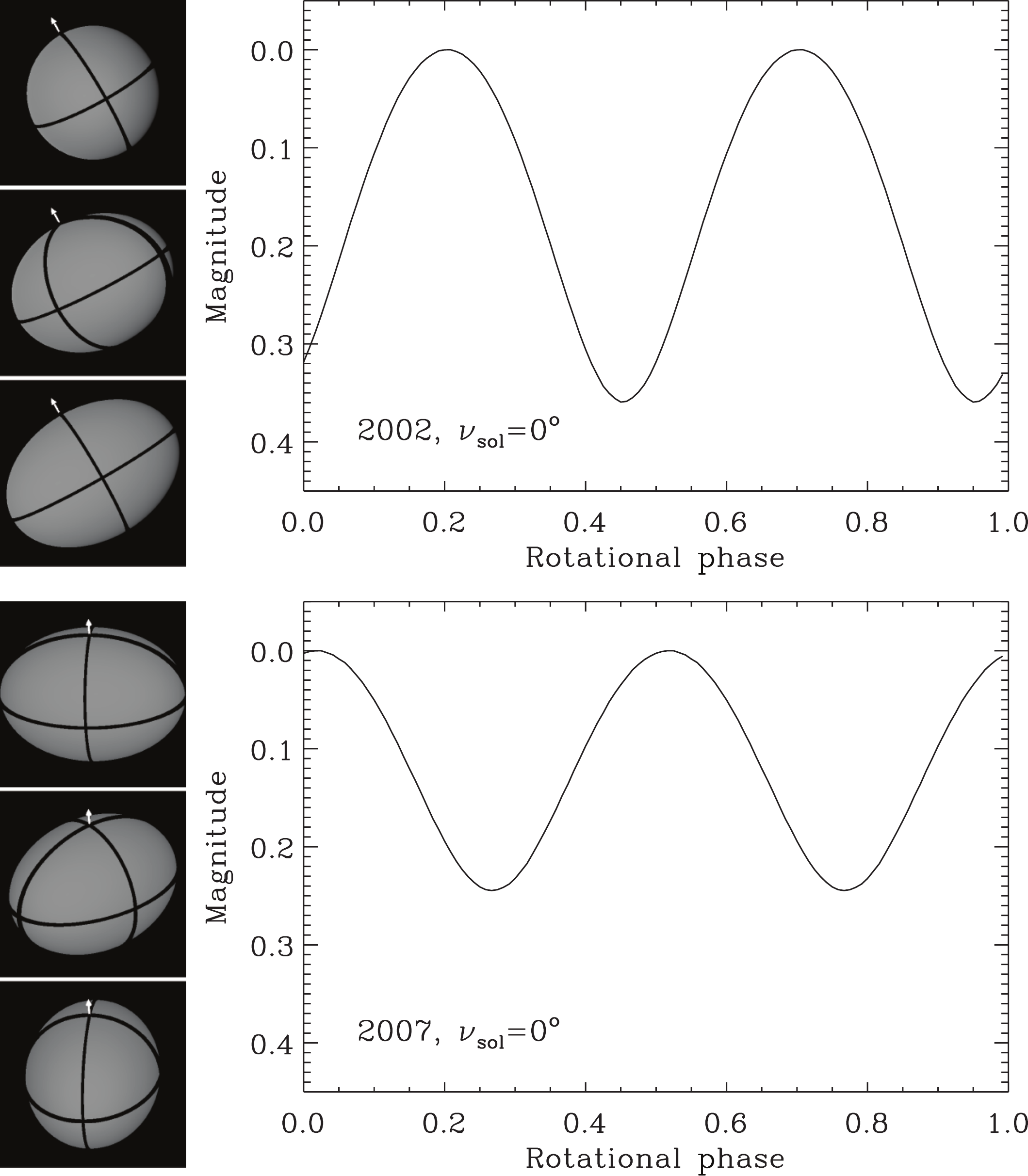}
\caption{\small 
Prolate ellipsoidal representation of 133P rendered as seen from Earth during
the September 2002 (upper panel) and July 2007 (lower panel)
observing campaigns. The pole orientation is set
such that $\nu_\mathrm{sol}=0\degr$ and 
obliquity is assumed to be $\varepsilon=30\degr$. A white line segment indicates
the position and orientation of the rotational pole, and black equator and meridian lines are
drawn on 133P's idealised surface to guide the eye.  Cross-sections at various rotational
phases are displayed from
top to bottom along the left of each panel, with the corresponding model light curve plotted in the right
portion of each panel.
}
\label{ltcurve_nu0}
\end{figure}

\begin{figure}
\includegraphics[width=6.5in]{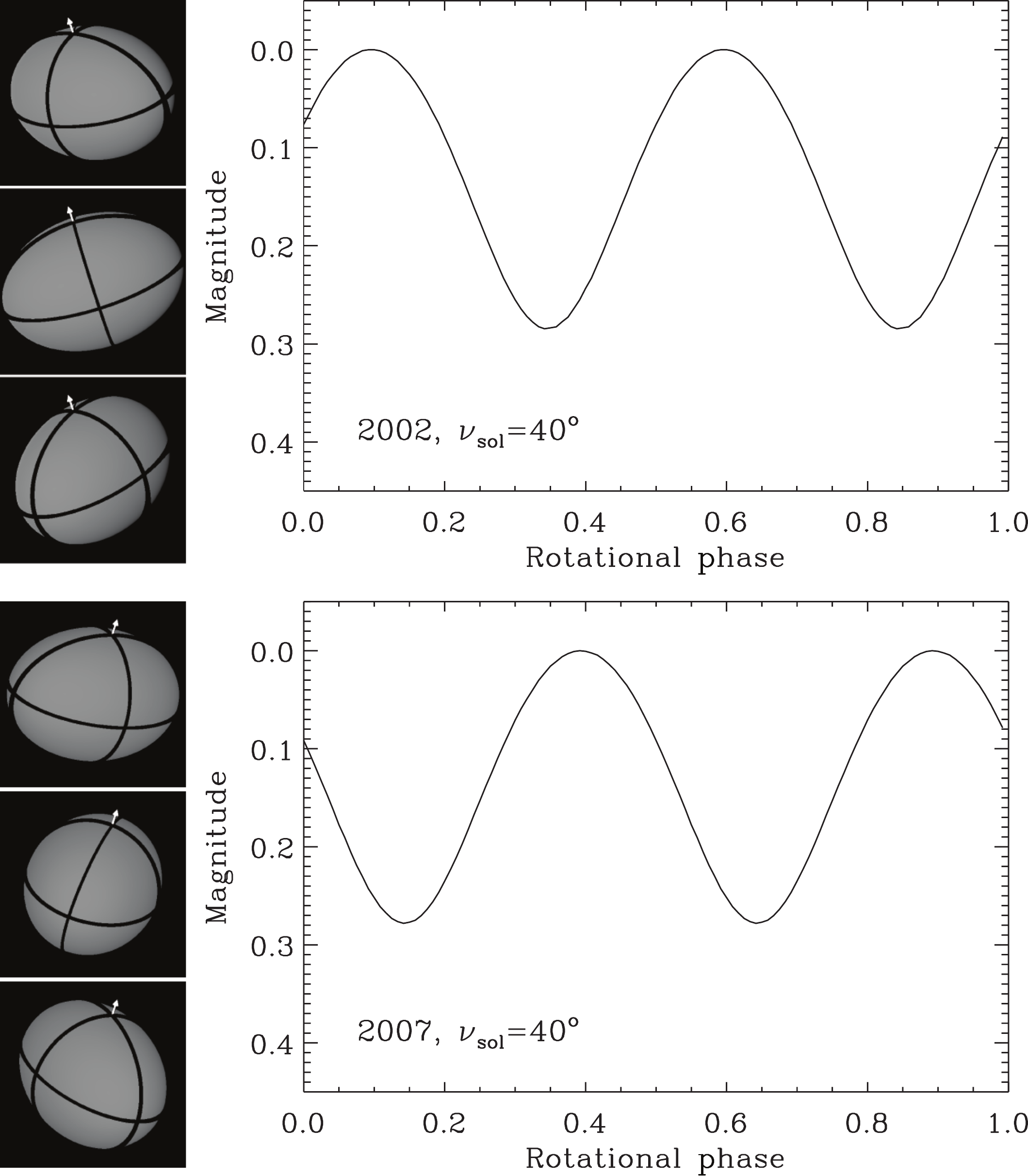}
\caption{\small 
Prolate ellipsoidal representation of 133P rendered as seen from Earth during
the September 2002 (upper panel) and July 2007 (lower panel)
observing campaigns. The pole orientation is set
such that $\nu_\mathrm{sol}=40\degr$ and 
obliquity is assumed to be $\varepsilon=30\degr$. A white line segment indicates
the position and orientation of the rotational pole, and black equator and meridian lines are
drawn on 133P's idealised surface to guide the eye.  Cross-sections at various rotational
phases are displayed from
top to bottom along the left of each panel, with the corresponding model light curve plotted in the right
portion of each panel.
}
\label{ltcurve_nu40}
\end{figure}

\begin{figure}
\includegraphics[width=6.5in]{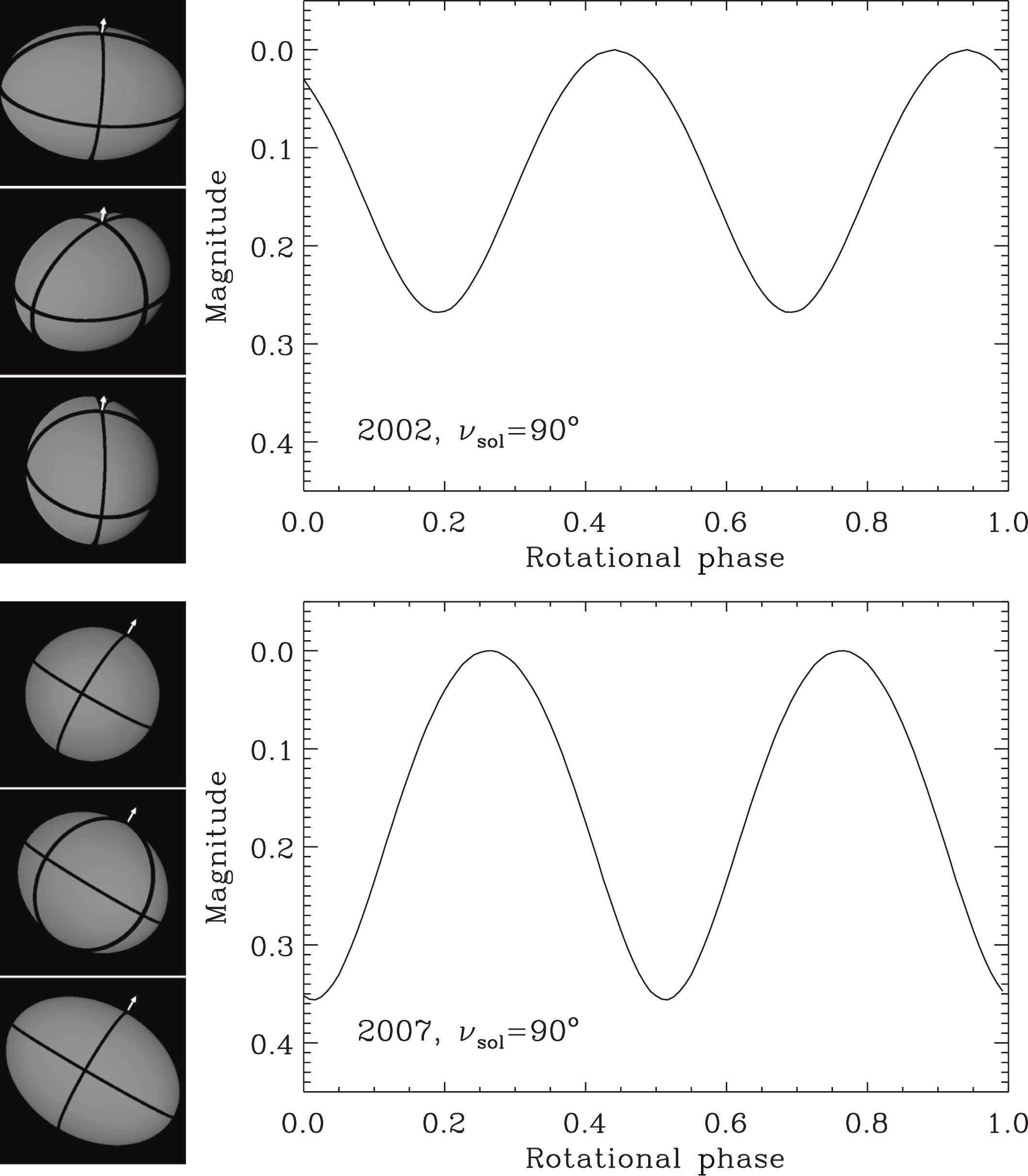}
\caption{\small 
Prolate ellipsoidal representation of 133P rendered as seen from Earth during
the September 2002 (upper panel) and July 2007 (lower panel)
observing campaigns. The pole orientation is set
such that $\nu_\mathrm{sol}=90\degr$ and 
obliquity is assumed to be $\varepsilon=30\degr$. A white line segment indicates
the position and orientation of the rotational pole, and black equator and meridian lines are
drawn on 133P's idealised surface to guide the eye.  Cross-sections at various rotational
phases are displayed from
top to bottom along the left of each panel, with the corresponding model light curve plotted in the right
portion of each panel.
}
\label{ltcurve_nu90}
\end{figure}

The seasonal heating hypothesis implies that 133P is at solstice when
close to perihelion, i.e. has a true anomaly at solstice of $\nu_\mathrm{sol}\approx0\degr$,
and also requires that the object have non-zero obliquity ($\varepsilon\neq0\degr$).
In principle, $\nu_\mathrm{sol}$ could potentially have any value from $\nu_\mathrm{sol}\approx0\degr$
to $\nu_\mathrm{sol}\approx45\degr$, since the
temperature of the hemisphere where 133P's active site is located will begin to rise
due to solar heating before the spin axis direction is actually aligned
with the Sun.  The seasonal heating hypothesis is inconsistent, however,
for pole orientations for which $\nu_\mathrm{sol}\approx90\degr$, or $\varepsilon=0\degr$.
We simulate the lightcurve behaviour of 133P for
$\nu_\mathrm{sol}=0\degr$, $\nu_\mathrm{sol}=40\degr$ and
$\nu_\mathrm{sol}=90\degr$. The first and third pole orientations are limiting cases that are
consistent and inconsistent with the seasonal heating hypothesis, respectively.
The intermediate geometry, in which solstice is reached approximately half-way through the active portion
of the orbit, is meant to test how sensitive we are to the exact longitude of
the pole.  Because we assume zero orbital inclination, each case sets the
ecliptic longitude of pole, and the ecliptic latitude is defined by the choice
of obliquity. We simulate obliquities of $\varepsilon=0\degr$,
$\varepsilon=10\degr$, $\varepsilon=20\degr$ and $\varepsilon=30\degr$.
Only $\varepsilon=0\degr$ is inconsistent with the seasonal
hypothesis. Rendered samples of 133P, where we assume $\varepsilon=30\degr$,
are shown in Figures~\ref{ltcurve_nu0}, \ref{ltcurve_nu40} and \ref{ltcurve_nu90}.

\begin{figure}
\includegraphics[width=6.5in]{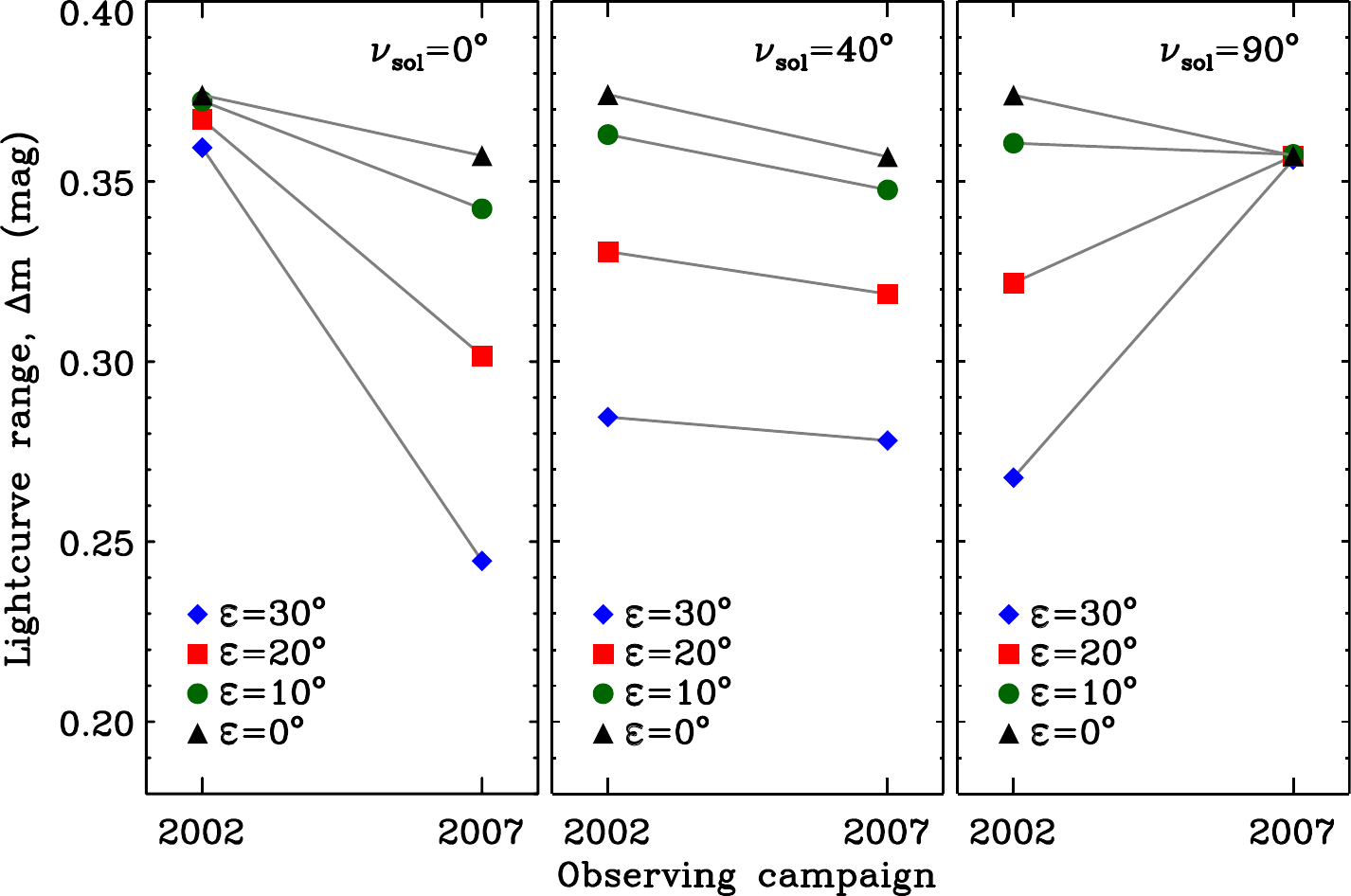}
\caption{\small
Expected photometric ranges for 133P during the 2002 September and the 2007 July
observing campaigns for pole solutions where (a) $\nu_\mathrm{sol}=0\degr$,
(b) $\nu_\mathrm{sol}=40\degr$, and (c) $\nu_\mathrm{sol}=90\degr$.  Each panel
shows predicted photometric ranges for obliquities $\varepsilon=0\degr$,
$\varepsilon=10\degr$, $\varepsilon=20\degr$, and $\varepsilon=30\degr$,
as labeled.  The seasonal heating hypothesis is consistent with the pole
orientations represented in (a) and (b), but not (c).
}
\label{rangevsgeom}
\end{figure}

Figure~\ref{rangevsgeom} shows the expected photometric range in 2002 September and
2007 July for each pole orientation. As expected, the photometric
range changes in opposite directions for $\nu_\mathrm{sol}=0\degr$ and
$\nu_\mathrm{sol}=90\degr$, whereas the intermediate pole orientation ($\nu_\mathrm{sol}=40\degr$)
produces
only a small change between the two epochs. The absolute value of $\Delta m_R$
in the figure depends on the assumed axis ratio and is
unimportant in this analysis, in which we are primarily concerned with relative changes.
The key feature is the variation of the range
between the two epochs.  The two possible scenarios indicated by the
data ({\it cf}. Fig.~\ref{ltcurves0207}) are where (a) both the 2002 and 2007 photometric
ranges are similar ($\Delta m_R\sim0.35$ mag), or (b) the 2002 photometric
range ($\Delta m_R\sim 0.35$ mag) is larger than the 2007 range ($\Delta
m_R\sim0.25$ mag). Inspection of Figure~\ref{rangevsgeom} shows that the first scenario is
consistent with low obliquity ($\varepsilon\lesssim10\degr$) and any of the considered pole orientations.
The second scenario is only consistent with a solstice around
$\nu_\mathrm{sol}=0\degr$ and significant obliquity ($\varepsilon\gtrsim20\degr$). Both
scenarios rule out a pole orientation where $\nu_\mathrm{sol}=90\degr$ if there is
also significant obliquity.  

\begin{figure}
\includegraphics[width=6.5in]{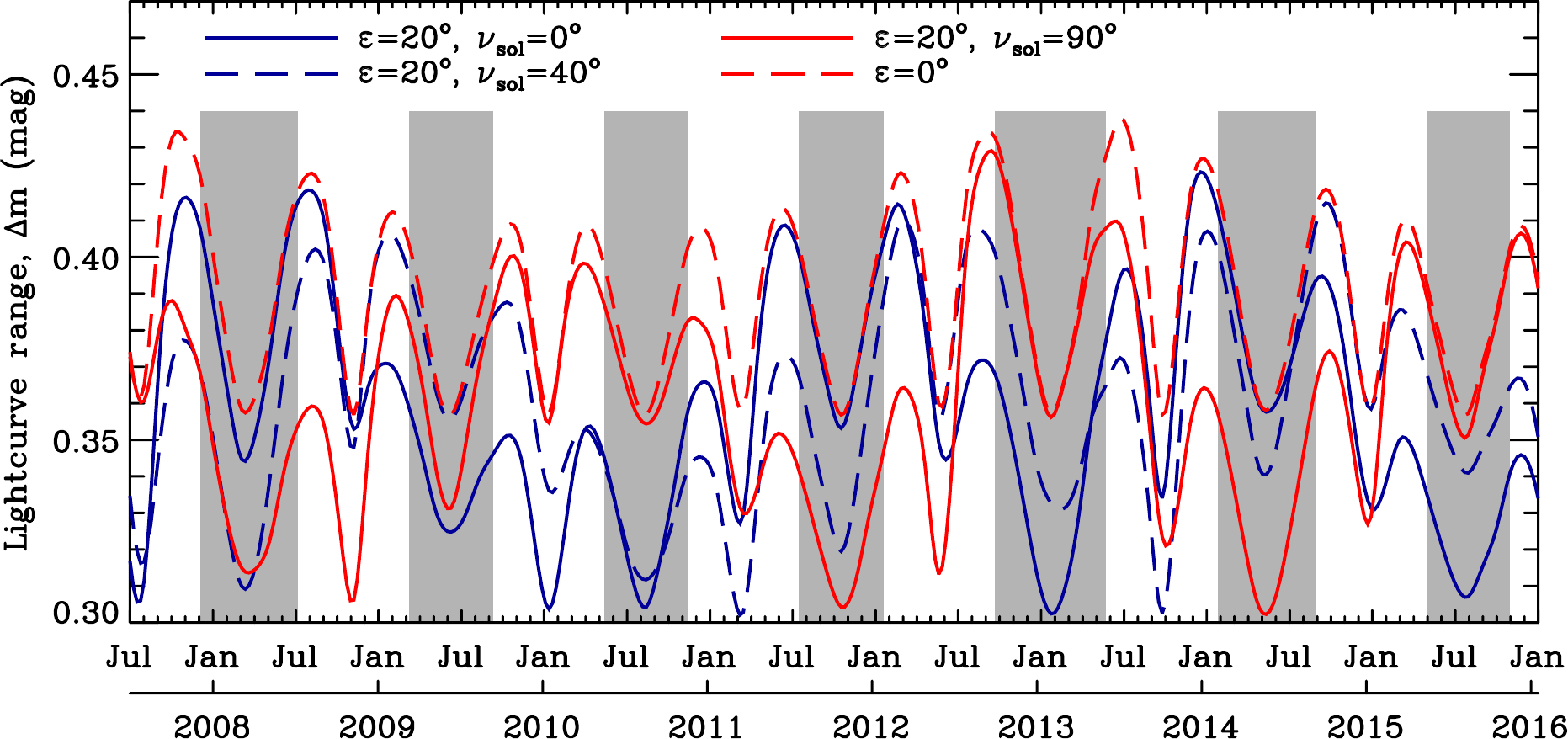}
\caption{\small Photometric range for 133P predicted by four different pole solutions
for the period between its 2007 July perihelion passage and 2016 January aphelion
passage. Dark lines correspond to solutions that are consistent with the
seasonal heating hypothesis, while light lines correspond to solutions that are
inconsistent with that hypothesis. Grey areas indicate times when 133P is at
solar elongations less than 80\degr, i.e., approximately when it is
observable for fewer than 4 hours in a single night.
}
\label{rangevsorbit}
\end{figure}

Clearly, additional and more complete lightcurve observations at different
points in 133P's orbit are needed to clarify how 133P's photometric
range varies with orbit position, constrain the object's pole orientation, and
determine whether the seasonal heating hypothesis remains plausible.
Given our current data, we can neither confirm nor reject the plausibility of
seasonal activity modulation as described by \citet{hsi04}.  While the pattern
of activity of 133P along its orbit appears consistent with the seasonal heating
hypothesis, the discovery of an incompatible pole solution could indicate that
activity is in fact modulated by factors other than obliquity, {\it e.g.},
shadowing of the active site by crater walls or other local topographic features.
In Figure~\ref{rangevsorbit}, we use
our model to forecast the photometric range behaviour of 133P over 1.5
orbits from its perihelion passage in 2007 August and to its
aphelion passage in 2016 January. We plot solutions for four pole positions,
two consistent with the seasonal heating hypothesis ( $\varepsilon=20\degr$, $\nu_\mathrm{sol}=0\degr$, and
$\varepsilon=20\degr$, $\nu_\mathrm{sol}=40\degr$) and two inconsistent with
that hypothesis
($\varepsilon=20\degr$, $\nu_\mathrm{sol}=90\degr$, and $\varepsilon=0\degr$).
The observability of 133P during this period is also indicated in the figure,
and should assist in planning observations
that are best-suited for discriminating between the various pole orientations
that we consider here.

\subsection{The Dust Trail Revisited\label{dusttrail}}

To produce deep composite images from our 2007 NTT data, we use linear interpolation
to shift the multiple images
obtained in each filter to align the photocenters of the nucleus in each
image, and sum the resulting shifted images.  
To measure the surface brightness profiles of the dust trail in these
composite images, we then rotate the images to make the trail horizontal in
the image frames, and measure the net flux in rectangular apertures placed
along the length of the trail \citep[{\it cf}.][]{hsi04}.
The dimensions of these equally-sized
apertures are set to lengths (along the direction of the trail) of 5 pixels,
and widths (perpendicular to the trail) of 6 pixels (approximately equal to
the FWHM of the trail cross-section on each night).  The net fluxes in these
apertures are then converted to net fluxes per linear arcsec (as measured
along the length of the trail) and normalised with respect to the total net
flux of the nucleus.

\begin{figure}
\includegraphics[width=6.5in]{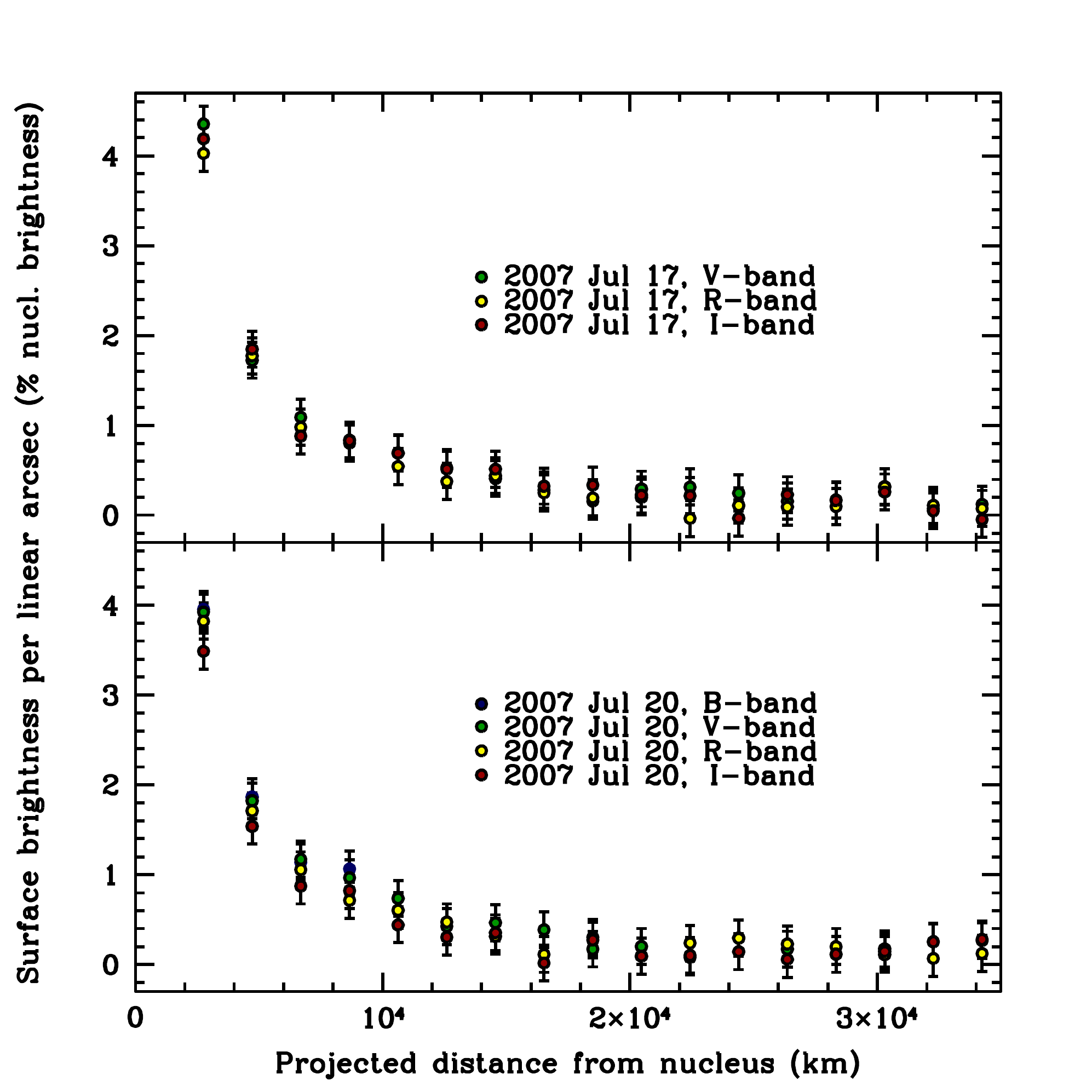}
\caption{\small Surface brightness profiles of the 133P dust trail, normalised to nucleus brightness,
  for composite images from observations made on 2007 Jul 17 (900s effective exposure time in each filter)
  and 2007 Jul 20 (360s effective exposure time in $B$, and 540 s effective exposure time in $V$, $R$, and $I$)
  using the NTT.  $B$-band data for 2007 Jul 20 have been truncated beyond 10$^4$~km due to contamination
  from nearby field stars.
}
\label{trailcolors07}
\end{figure}

We plot the resulting surface brightness profiles for
both 2007 Jul 17 and 2007 Jul 20 in Figure~\ref{trailcolors07}.  From these
plots, we see that the trail profile does not change significantly
between the two nights.  We also note that there are minimal
differences in the profiles of the trail as observed in different
filters, indicating that the colours of the dust along the trail are consistently
similar to those of the nucleus.  To quantify this observation, we measure the
surface brightness of the trail as observed on 2007 July 20 in each filter
in a single aperture approximately 5 arcsec (15 pixels, or $\sim$6000~km) in length
and 1 arcsec (3 pixels or $\sim$1200~km) in width placed along the trail.
Seeking to minimise the effect of the nucleus on our trail photometry, we place
the nearest edge of this aperture $\sim$3.0~arcsec from the nucleus photocentre.
We find surface brightnesses of $\Sigma_B=24.88\pm0.17$~mag~arcsec$^{-2}$,
$\Sigma_V=24.35\pm0.05$~mag~arcsec$^{-2}$, $\Sigma_R=24.04\pm0.05$~mag~arcsec$^{-2}$,
and $\Sigma_I=23.70\pm0.07$~mag~arcsec$^{-2}$, giving colours of
$B-V=0.53\pm0.18$~mag~arcsec$^{-2}$, $V-R=0.31\pm0.07$~mag~arcsec$^{-2}$,
and $R-I=0.34\pm0.09$~mag~arcsec$^{-2}$, consistent with the colours of the nucleus
found in \S\ref{ltcurverevisit}.

We also wish to know how trail morphology changes between 133P's 2002 and 2007
active episodes.  The most obvious difference between the two observing epochs
is that the dust trail of 133P is significantly shorter in our 2007 data
than in 2002 (despite composite images from each epoch being of
approximately equivalent effective exposure time),
extending only $\sim$30 arcsec from the nucleus in
2007 observations, compared to nearly 3 arcmin in 2002 \citep{hsi04}.
In terms of trail width, the observed mean FWHM of the trail on 2002 Sep 07
over the first 10~arcsec of the trail, as measured from the edge of the nucleus's
seeing disc (taken to be 2.5$\times$ the FWHM seeing), was measured to be
$\theta_o=1\farcs3$.  This observed value
corresponds to an intrinsic FWHM of $\theta_i=0\farcs9$
($\sim$1300~km in the plane of the sky), which is computed using
\begin{equation}
\theta_i = (\theta_o^2 - \theta_s^2)^{1/2}
\label{intrinsicwidth}
\end{equation}
where the FWHM seeing was $\theta_s=0\farcs9$ on 2002 Sep 07.
For comparison, the observed FWHM of the trail on 2007 July 17 was
$\theta_o=1\farcs9$,
corresponding to $\theta_i=1\farcs3$
($\sim$1500~km in the plane of the sky),
where $\theta_s=1\farcs4$.
Given that viewing geometries (parametrized by out-of-plane viewing
angles, $\alpha_{pl}$) in 2002 and 2007 were comparable,
we therefore find that the computed intrinsic width of the dust trail
is approximately equal in both our 2002 and 2007 observations.  
As \citet{hsi04} found the
primary factor controlling 133P's trail width to be particle
ejection velocity,
this result suggests that sublimation
took place with comparable intensity in both 2002 and 2007.

\begin{figure}
\includegraphics[width=6.5in]{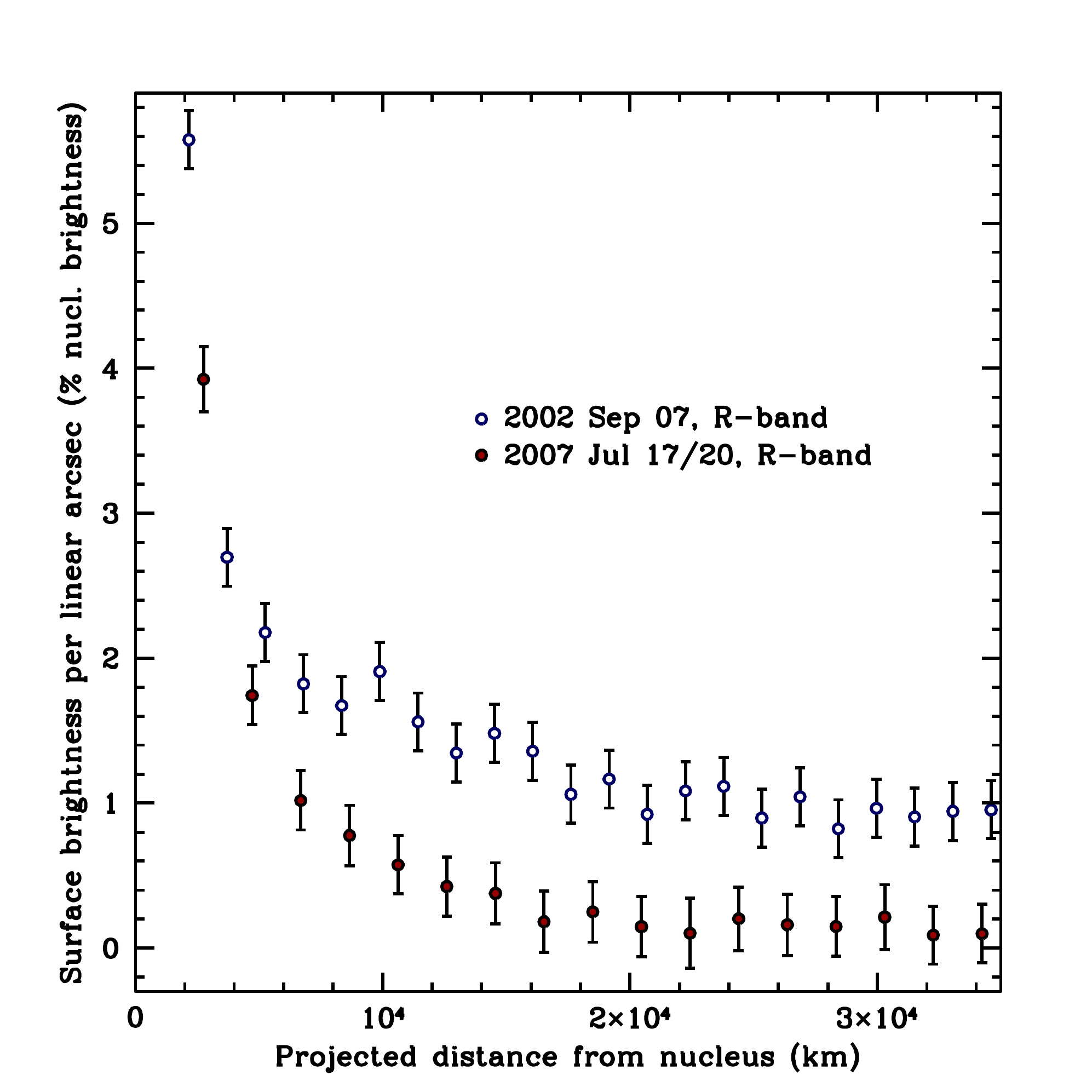}
\caption{\small Surface brightness profiles of the 133P dust trail, normalised to nucleus brightness,
  for composite images from observations made on 2002 Sep 07 (3900 s effective exposure time; equivalent to
  1475 s on the NTT) using the UH 2.2-m telescope, and 2007 Jul 17 (900 s effective exposure time)
  and 2007 Jul 20 (540 s effective exposure time) using the NTT.
}
\label{trailcomparison}
\end{figure}

In order to further compare 133P's activity level in 2002 and 2007, we measure
the profile of 133P's trail in $R$-band data from 2002 September 07 using the
procedure described above, i.e., using rectangular apertures
placed along the length of the trail with lengths of 5 pixels and widths
of 6 pixels each.  We then compare the resulting profile
to the mean $R$-band trail profile from 2007
(Fig.~\ref{trailcomparison}), finding that the
trail is noticeably weaker in 2007 than it was in 2002.  

The difference in trail strength in 2002 and 2007 could be due to
several factors.  The simplest explanation is that the activity was
actually weaker in 2007 due to depletion of exposed volatile
material on 133P by the previous outburst.  This explanation, however, is at odds with
our findings of comparable dust ejection velocities for the two observing
epochs (above), and comparable dust enhancement of the nucleus brightness
(\S\ref{activitydetection}).  A more likely explanation is that by the
time our 2007 NTT observations were made, 133P was no more than 4 months into
its current active phase, whereas it had been
active for about a year by the time it was observed on
2002 September 07.  Thus, 133P may simply have not yet reached its peak
level of activity by the time we observed it with the NTT in 2007.
The position of 133P near opposition on 2007 July 17
and 20 also meant that a dust trail pointed in the antisolar direction would be
highly projected in the sky, which would additionally explain why
the trail appeared to be so much shorter in 2007 than in 2002.

\section{SUMMARY}

Key results are as follows:
\begin{enumerate}
\item{Monitoring observations of 133P show no evidence of activity from
      UT 2003 September 22 through UT 2007 March 21.  This result is consistent with
      the seasonal activity modulation hypothesis proposed by \citet{hsi04}
      which predicted that, following its 2002 outburst, 133P should remain
      inactive until approximately late 2007.
}
\item{A recomputation of 133P's phase function parameters using inactive data
      yielded the new IAU phase function parameters of $H_R=15.49\pm0.05$~mag
      and $G_R=0.04\pm0.05$, and linear phase function parameters of
      $m_R(1,1,0)=15.80\pm0.05$~mag and $\beta=0.041\pm0.005$~mag~deg$^{-1}$.
      While these new values for $G_R$ and $\beta$ are similar to the values
      computed by \citet{hsi04}, the values for $H_R$ and $m_R(1,1,0)$ computed
      here are $\sim$0.2~mag fainter than previously derived values, a discrepancy
      we attribute to previously undetected dust contamination.
}
\item{Comparison of 133P's newly-computed IAU phase function with rotationally-averaged
      magnitudes found during its 2002 active outburst reveals the
      presence of unresolved coma with a dust scattering surface area on the order
      of $\sim$0.20 of the nucleus cross-section.  Similarly, unresolved coma and trail material
      on the
      order of $\sim$0.25 of the nucleus cross-section is found in images taken on
      2007 July 17 and 20, increasing to $\sim$0.65 of the nucleus cross-section
      in August and September as the dust trail appears to become projected almost directly
      behind the nucleus as viewed from Earth.
}
\item{From NTT observations obtained in 2007, we find mean nucleus colours of
      $B-V=0.65\pm0.03$, $V-R=0.36\pm0.01$, and $R-I=0.32\pm0.01$, and no
      evidence of colour inhomogeneities on 133P's surface (though we hypothesize
      that inhomogeneities will be more effectively searched for between late 2011
      and early 2012).  Additionally, we find
      from the same observations that the
      dust trail shares approximately the same colours as the nucleus.
}
\item{Examination of coma-corrected lightcurve data for 133P from 2002 and 2007 indicates a possible
      reduction of photometric range from $\Delta m_R\sim0.35$~mag in 2002
      to $\Delta m_R\sim0.25$~mag in 2007, though this result is inconclusive due to
      incomplete sampling of the lightcurve in 2007.  Additional observations will
      be needed to determine how 133P's photometric range actually varies with
      orbital position and what implications these variations have for constraining
      the object's pole orientation.  Our present
      constraints on pole orientation do not currently
      permit us to confirm or reject obliquity-related seasonal activity modulation as a plausible
      mechanism for explaining 133P's active behaviour.
}
\item{While 133P's dust trail appears shorter and weaker in 2007 data as compared to
      2002 data, other measures of activity strength (dust ejection velocity and
      dust contamination of nucleus photometry) during the two outburst events
      are found to remain roughly constant.  We suggest that the weaker trail observed
      in 2007 could simply be due to the fact that observations
      were made at an earlier stage in 133P's active phase than in 2002,
      and find that there is no conclusive
      evidence of any substantial decrease in activity strength between 2002 and 2007.
}

\end{enumerate}

\section*{Acknowledgements}
We thank John Dvorak, Dave Brennen, Dan Birchall, Ian Renaud-Kim, and Jon Archambeau at the UH 2.2-m,
Greg Wirth, Cynthia Wilburn, and Gary Punawai at Keck, Michelle Buxton and various queue observers at NOAO,
and Leonardo Gallegos at the NTT for their
assistance with our observations, and Matthew Knight for a prompt and helpful review.
We appreciate support of
this work through STFC fellowship grant ST/F011016/1 to HHH,
NASA planetary astronomy grants to DJ and SCL, a Royal Society
Newton Fellowship grant to PL,
the National Optical Astronomy
Observatory, and the European Southern Observatory.

\begin{table}
\begin{minipage}[t]{\columnwidth}
\caption{Observations of 133P/Elst-Pizarro}
\label{obs_elstpiz}
\centering
\renewcommand{\footnoterule}{}
\begin{tabular}{llcrccccrccrr}
\hline\hline
  UT Date
   & Tel.\footnote{Telescope used (UH2.2: University of Hawaii 2.2-m
                   telescope; Keck: Keck I 10-m telescope;
		   NTT: 3.58-m New Technology Telescope;
		   CT1.3: SMARTS 1.3-m telescope at Cerro Tololo )}
   & $N$\footnote{Number of images}
   & $t$\footnote{Total effective exposure time in seconds}
   & Filters
   & $m_R$\footnote{Observed mean $R$-band magnitude of nucleus}
   & $m_{mid}$\footnote{Estimated $R$-band magnitude at midpoint of full photometric range (assumed to be 0.40~mag) of rotational lightcurve}
   & $\theta_s$\footnote{FWHM seeing in arcsec}
   & $\nu$\footnote{True anomaly in degrees}
   & $R$\footnote{Median heliocentric distance in AU}
   & $\Delta$\footnote{Median geocentric distance in AU}
   & $\alpha$\footnote{Solar phase angle in degrees}
   & $\alpha_{pl}$\footnote{Orbit plane angle (between the observer and object orbit
                   plane as seen from the object) in degrees} \\
\hline
2001 Nov 23 
  & {\it Perihelion} & ... & ... & ... & ... & ... & ...
  & 0.0 & 2.64 & 3.34 & 13.5 & 0.4 \\
2002 Aug 19\footnote{\citet{hsi04}} & UH2.2  &  6 &  2500 & $R$ & 20.05$\pm$0.02 & 20.05$\pm$0.10 & 1.1 &  63.3 & 2.86 & 2.05 & 14.5 & --0.2 \\
2002 Sep 07$^{k}$ & UH2.2  & 14 &  4200 & $R$    & 19.71$\pm$0.01 & 19.70$\pm$0.05 & 0.8 &  67.2 & 2.89 & 1.94 &  8.2 &   0.1 \\
2002 Sep 08$^{k}$ & UH2.2  &  6 &  1800 & $R$    & 19.71$\pm$0.02 & 19.70$\pm$0.05 & 0.8 &  67.4 & 2.89 & 1.93 &  7.8 &   0.1 \\
2002 Sep 09$^{k}$ & UH2.2  & 14 &  4200 & $BVRI$ & 19.63$\pm$0.05 & 19.70$\pm$0.05 & 1.2 &  67.6 & 2.89 & 1.93 &  7.6 &   0.1 \\
2002 Nov 05$^{k}$ & UH2.2  &  5 &  1500 & $R$    & 20.28$\pm$0.03 & 20.25$\pm$0.05 & 0.6 &  79.1 & 2.98 & 2.18 & 13.3 &   0.6 \\
2002 Nov 06$^{k}$ & UH2.2  &  5 &  1500 & $R$    & 20.13$\pm$0.03 & 20.25$\pm$0.05 & 0.9 &  79.3 & 2.98 & 2.19 & 13.6 &   0.6 \\
2002 Nov 07$^{k}$ & UH2.2  &  4 &  1200 & $R$    & 20.28$\pm$0.03 & 20.25$\pm$0.05 & 0.9 &  79.4 & 2.98 & 2.21 & 13.8 &   0.6 \\
2002 Dec 27$^{k}$ & UH2.2  & 20 &  6000 & $BVRI$ & 21.14$\pm$0.04 & 21.20$\pm$0.05 & 1.1 &  88.9 & 3.06 & 2.93 & 18.7 &   0.4 \\
2002 Dec 28$^{k}$ & UH2.2  & 11 &  3300 & $BVRI$ & 21.19$\pm$0.05 & 21.20$\pm$0.05 & 1.2 &  89.1 & 3.06 & 2.95 & 18.7 &   0.4 \\
2003 Sep 22$^{k}$ & Keck   &  9 &   900 & $BVR$  & 21.71$\pm$0.06 & 21.71$\pm$0.15 & 0.6 & 133.7 & 3.46 & 3.19 & 16.8 &   0.1 \\
2003 Dec 13      & UH2.2  &  3 &   900 & $R$    & 20.45$\pm$0.04 & 20.45$\pm$0.20 & 1.3 & 144.3 & 3.54 & 2.56 &  1.5 &   0.5 \\
2003 Dec 15      & UH2.2  &  3 &   900 & $R$    & 20.70$\pm$0.06 & 20.70$\pm$0.20 & 1.2 & 144.5 & 3.55 & 2.57 &  2.1 &   0.5 \\
2004 Feb 16      & Keck   &  1 &   300 & $R$    & 21.69$\pm$0.06 & 21.69$\pm$0.20 & 1.2 & 153.1 & 3.60 & 3.24 & 15.5 &   0.1 \\
2004 Sep 11      & {\it Aphelion} & ... & ... & ... & ... & ... & ...
  & 180.0 & 3.68 & 4.33 & 11.1 & 0.0 \\
2004 Oct 10      & Keck   &  2 &   200 & $R$    & 22.15$\pm$0.04 & 22.15$\pm$0.20 & 0.6 & 183.8 & 3.68 & 3.96 & 14.4 &   0.2 \\
2005 Jan 16      & UH2.2  &  2 &   600 & $R$    & 20.99$\pm$0.05 & 20.99$\pm$0.20 & 0.9 & 196.4 & 3.65 & 2.70 &  4.7 &   0.3 \\
2005 Apr 10      & UH2.2  &  4 &  1200 & $R$    & 21.86$\pm$0.12 & 21.86$\pm$0.15 & 1.1 & 207.5 & 3.60 & 3.22 & 15.7 & --0.3 \\
2005 May 27      & UH2.2  &  4 &  1200 & $R$    & 22.03$\pm$0.20 & 22.03$\pm$0.20 & 1.0 & 213.8 & 3.56 & 3.85 & 15.0 & --0.4 \\
2005 May 28      & UH2.2  &  4 &  1200 & $R$    & 22.08$\pm$0.21 & 22.08$\pm$0.21 & 1.6 & 213.9 & 3.56 & 3.87 & 15.0 & --0.4 \\
2005 Dec 27      & UH2.2  & 12 &  3600 & $R$    & 21.70$\pm$0.06 & 21.72$\pm$0.10 & 0.8 & 245.3 & 3.30 & 3.31 & 17.1 &   0.4 \\
2006 Apr 23      & UH2.2  &  4 &  1200 & $R$    & 20.09$\pm$0.04 & 20.09$\pm$0.15 & 0.9 & 265.1 & 3.12 & 2.16 &  7.1 & --0.5 \\
2006 May 22      & UH2.2  &  4 &  1200 & $R$    & 20.84$\pm$0.05 & 20.84$\pm$0.15 & 1.0 & 270.4 & 3.07 & 2.36 & 15.4 & --0.6 \\
2006 May 23      & UH2.2  &  2 &   600 & $R$    & 20.73$\pm$0.05 & 20.73$\pm$0.20 & 0.8 & 270.6 & 3.07 & 2.37 & 15.6 & --0.6 \\
2006 May 25      & UH2.2  &  2 &   600 & $R$    & 21.04$\pm$0.11 & 21.04$\pm$0.20 & 1.1 & 270.9 & 3.07 & 2.39 & 16.0 & --0.6 \\
2007 Mar 21      & UH2.2  &  4 &  1200 & $R$    & 21.15$\pm$0.11 & 21.15$\pm$0.15 & 1.1 & 335.5 & 2.68 & 2.81 & 20.7 & --0.2 \\
2007 May 19      & UH2.2  & 11 &  3300 & $R$    & 20.19$\pm$0.02 & 20.10$\pm$0.10 & 1.2 & 349.9 & 2.65 & 2.04 & 20.1 & --0.7 \\
2007 Jun 29 & {\it Perihelion} & ... & ... & ... & ... & ... & ...
  & 0.0 & 2.64 & 1.68 & 9.0 & --0.7 \\
2007 Jul 17 & NTT    & 41 &  2460 & $VRI$  & 18.59$\pm$0.02 & 18.65$\pm$0.05 & 1.4 &   4.5 & 2.64 & 1.63 &  1.7 & --0.6 \\
2007 Jul 20 & NTT    & 54 &  3240 & $BVRI$ & 18.53$\pm$0.01 & 18.55$\pm$0.05 & 1.1 &   5.2 & 2.64 & 1.63 &  0.6 & --0.6 \\
2007 Aug 18 & CT1.3  & 70 &  4200 & $R$    & 19.08$\pm$0.04 & 19.10$\pm$0.10 & 1.8 &  12.3 & 2.65 & 1.74 & 11.6 & --0.2 \\
2007 Sep 12 & CT1.3  & 69 &  4140 & $R$    & 19.63$\pm$0.05 & 19.60$\pm$0.10 & 1.4 &  18.4 & 2.66 & 1.97 & 18.4 &   0.1 \\
2008 Jul 01 & NTT    &  2 &   600 & $R$    & 21.42$\pm$0.21 & 21.42$\pm$0.21 & 1.2 &  82.8 & 3.01 & 3.27 & 18.1 & --0.4 \\
2010 Apr 20 & {\it Aphelion} & ... & ... & ... & ... & ... & ...
  & 180.0 & 3.68 & 3.93 & 14.7 & --0.3 \\
2013 Feb 09 & {\it Perihelion} & ... & ... & ... & ... & ... & ...
  &   0.0 & 2.64 & 3.55 &  7.5 & 0.1 \\
\hline
\end{tabular}
\end{minipage}
\end{table}

\begin{table}
\begin{minipage}[t]{\columnwidth}
\caption{Dust Contribution to 133P Photometry}
\label{dustcontrib}
\centering
\renewcommand{\footnoterule}{}
\begin{tabular}{lcccccrc}
\hline\hline
 UT Date
  & Obs.\footnote{Type of observations (LC: observations of complete lightcurve; SS: snapshot observations at unknown rotational phase)}
  & $m_{mid}$\footnote{Estimated $R$-band magnitude at midpoint of full photometric range (assumed to be 0.40~mag) of rotational lightcurve}
  & $m_{exp}$\footnote{Expected $R$-band magnitude using best-fit IAU phase function}
  & $A_{dust}/A_{nucl}$\footnote{Inferred ratio of scattering surface area of dust to nucleus scattering surface area}
  & $A_{dust}$\footnote{Inferred scattering surface area of dust, in $10^6$ m$^2$, using $A_{nucl}=1.13\times10^7$~m$^2$}
  & $M_{dust}$\footnote{Estimated dust mass, in $10^4$~kg, assuming 10~$\mu$m-radius grains and $\rho=1300$~kg~m$^{-3}$}
  & $Af\rho$\footnote{Dust contribution inside $4\farcs0$ aperture, as parametrized by \citet{ahe84}} \\
\hline
2002 Aug 19    & LC & 20.05$\pm$0.10 & 20.26$\pm$0.05 & 0.21$\pm$0.08 & 2.4$\pm$0.9 &  4.2$\pm$1.6 & 17.4$\pm$0.8 \\
2002 Sep 07-09 & LC & 19.70$\pm$0.05 & 19.88$\pm$0.05 & 0.18$\pm$0.08 & 2.0$\pm$0.9 &  3.5$\pm$1.6 & 17.5$\pm$0.8 \\
2002 Nov 05-07 & LC & 20.25$\pm$0.05 & 20.43$\pm$0.05 & 0.18$\pm$0.08 & 2.0$\pm$0.9 &  3.5$\pm$1.6 & 16.3$\pm$0.8 \\
2002 Dec 27-28 & LC & 21.20$\pm$0.05 & 21.35$\pm$0.05 & 0.20$\pm$0.08 & 2.3$\pm$0.9 &  4.0$\pm$1.6 & 12.1$\pm$0.6 \\
2007 May 19    & SS & 20.10$\pm$0.10 & 20.30$\pm$0.05 & 0.20$\pm$0.13 & 2.3$\pm$1.5 &  4.0$\pm$2.6 & 17.4$\pm$1.6 \\
2007 Jul 17    & LC & 18.65$\pm$0.05 & 18.90$\pm$0.05 & 0.26$\pm$0.08 & 2.9$\pm$1.0 &  5.1$\pm$1.7 & 22.7$\pm$1.0 \\
2007 Jul 20    & LC & 18.55$\pm$0.05 & 18.79$\pm$0.05 & 0.25$\pm$0.08 & 2.8$\pm$1.0 &  4.9$\pm$1.6 & 22.3$\pm$1.0 \\
2007 Aug 18    & SS & 19.10$\pm$0.10 & 19.62$\pm$0.05 & 0.61$\pm$0.18 & 6.9$\pm$2.0 & 12.0$\pm$3.4 & 27.3$\pm$2.5 \\
2007 Sep 12    & SS & 19.60$\pm$0.10 & 20.17$\pm$0.05 & 0.69$\pm$0.18 & 7.8$\pm$2.1 & 13.5$\pm$3.6 & 25.3$\pm$2.3 \\
\hline
\end{tabular}
\end{minipage}
\end{table}

\clearpage

\bsp

\label{lastpage}

\end{document}